\RequirePackage{fix-cm}
\documentclass[natbib,smallextended]{svjour3}       
\smartqed  
\usepackage{graphicx}

\usepackage{amssymb}
\usepackage{amsmath}
\usepackage{bm}
\usepackage{subfigure}
\usepackage{tabularx} 
\usepackage{tabulary}

\graphicspath{}

\journalname{Space Science Reviews}
\newcommand{\figuremacroM}[6]{
	\begin{figure}[htbp!]
    	\centering
    	\subfigure[ ]
   	 {
        	\includegraphics[width=2.2in]{#1}
        	\label{fig:#1}
    	}
    	\subfigure[ ]
    	{
        	\includegraphics[width=2.2in]{#2}
        	\label{fig:#2}
    	}
	\\
    	\subfigure[ ] 
	{
        	\includegraphics[width=2.2in]{#3}
        	\label{fig:#3}
    	}
    	\caption[#5]{\textbf{#5} - #6}
    	\label{fig:#4}
	\end{figure}
	}

\newcommand{\figuremacroW}[4]{
	\begin{figure}[htbp!]
		\centering
		\includegraphics[width=#4\textwidth]{#1}
		\caption[#2]{\textbf{#2} - #3}
		\label{fig:#1}
	\end{figure}
}

\newcommand{\figuremacroTwo}[5]{
	\begin{figure}[htbp!]
    	\centering
   	 {
        	\includegraphics[height=2.5in]{#1}
        	\label{fig:#1}
    	}
    	{
        	\includegraphics[height=2.5in]{#2}
        	\label{fig:#2}
    	}
	\\
    	\caption[#4]{\textbf{#4} - #5}
    	\label{fig:#3}
	\end{figure}
	}

\newcommand{\figuremacroFive}[8]{
	\begin{figure}[htbp!]
    	\centering
    	\subfigure[ ]
   	 {
        	\includegraphics[width=1.65in]{#1}
        	\label{fig:#1}
    	}
    	\subfigure[ ]
    	{
        	\includegraphics[width=1.65in]{#2}
        	\label{fig:#2}
    	}
	\subfigure[ ]
    	{
        	\includegraphics[width=1.65in]{#3}
        	\label{fig:#3}
    	}
	\\
    	\subfigure[ ] 
	{
        	\includegraphics[width=1.65in]{#4}
        	\label{fig:#4}
    	}
	\subfigure[ ] 
	{
        	\includegraphics[width=1.65in]{#5}
        	\label{fig:#5}
    	}
    	\caption[#6]{\textbf{#7} - #8}
    	\label{fig:#6}
	\end{figure}
	}

\newcommand{\eg}{e.g., }

\newcommand{\ie}{i.e., }

\newcommand{\sect}[1]{Section \ref{s:#1}}

\newcommand{\fig}[1]{Fig.\ \ref{fig:#1}}

\newcommand{\Fig}[1]{Figure \ref{fig:#1}}

\newcommand{\tbl}[1]{Table \ref{t:#1}}

\newcommand{\tbls}[2]{Tables \ref{t:#1}--\ref{t:#2}}

\newcommand{\hide}[1]{} 

\begin{document}

\title{Evaluating the wind-induced mechanical noise on the InSight seismometers}


\author{Naomi Murdoch        \and
        David Mimoun \and
        Raphael F. Garcia \and
        William Rapin \and
        Taichi Kawamura \and
        Philippe Lognonn{\'e} \and
        Don Banfield \and
                William B. Banerdt 
}

\institute{N. Murdoch \at
              Institut Sup\'{e}rieur de l'A\'{e}ronautique et de l'Espace (ISAE-SUPAERO), Universit\'{e} de Toulouse, 31055 Toulouse Cedex 4, France \\
              \email{naomi.murdoch@isae.fr}           
           \and
          D. Mimoun \at
              Institut Sup\'{e}rieur de l'A\'{e}ronautique et de l'Espace (ISAE-SUPAERO), Universit\'{e} de Toulouse, 31055 Toulouse Cedex 4, France \\
           \and
           R. F. Garcia \at
              Institut Sup\'{e}rieur de l'A\'{e}ronautique et de l'Espace (ISAE-SUPAERO), Universit\'{e} de Toulouse, 31055 Toulouse Cedex 4, France \\
           \and
            W. Rapin \at
              L'Institut de Recherche en Astrophysique et PlaneŽtologie (IRAP), Universit\'{e} de Toulouse III Paul Sabatier, 31400 Toulouse, France \\
           \and
              T. Kawamura \at
             Institut de Physique du Globe de Paris, Paris, France \\
           \and
            P.  Lognonn{\'e}\at
             Institut de Physique du Globe de Paris, Paris, France \\
               \and
               D. Banfield\at
               Cornell Center for Astrophysics and Planetary Science, Cornell University, Ithaca, NY 14853, USA
			\and              
			W. Bruce Banerdt \at
             Jet Propulsion Laboratory, Pasadena, CA  91109, USA 
           }

\date{Submitted: Friday 17 June 2016, Re-submitted: Friday 14 October 2016}

\maketitle

\begin{abstract}
The SEIS (Seismic Experiment for Interior Structures) instrument onboard the InSight mission to Mars is the critical instrument for determining the interior structure of Mars, the current level of tectonic activity and the meteorite flux. Meeting the performance requirements of the SEIS instrument is vital to successfully achieve these mission objectives.  Here we analyse in-situ wind measurements from previous Mars space missions to understand the wind environment that we are likely to encounter on Mars, and then we use an elastic ground deformation model to evaluate the mechanical noise contributions on the SEIS instrument due to the interaction between the Martian winds and the InSight lander. Lander mechanical noise maps that will be used to select the best deployment site for SEIS once the InSight lander arrives on Mars are also presented. We find the lander mechanical noise may be a detectable signal on the InSight seismometers. However, for the baseline SEIS deployment position, the noise is expected to be below the total noise requirement $>$97\% of the time and is, therefore, not expected to endanger the InSight mission objectives.

\end{abstract}

\section{Introduction}
\label{sec:intro} 
The upcoming InSight mission, selected under the NASA Discovery programme for launch in 2018, will perform the first comprehensive surface-based geophysical investigation of Mars. InSight, which will land in Elysium Planitia, will help scientists to understand the formation and evolution of terrestrial planets and to determine the current level of tectonic activity and impact flux on Mars. SEIS (Seismic Experiment for Internal Structures) is the critical instrument for delineating the deep interior structure of Mars, including the thickness and structure of the crust, the composition and structure of the mantle, and the size of the core. 

SEIS consists of two independent, three-axis seismometers: an ultra-sensitive very broad band (VBB) oblique seismometer; and a miniature, short-period (SP) seismometer that provides partial measurement redundancy and extends the high-frequency measurement capability \citep{lognonne2015}. This combined broad band and short-period instrument architecture was also used on the Apollo Lunar Surface Experiments Package (ALSEP; with only one vertical SP axis) and for NetLander, ExoMars, and SELENE-2.  See \cite{lognonne2005}, \cite{lognonne2015} and \cite{lognonne2007} for general reviews on past planetary seismology missions and projects. For InSight, both instruments are mounted on the precision levelling structure (LVL) together with their respective signal preamplifier stages. The seismometers and the levelling structure will be deployed on the ground as an integrated package after arrival on Mars. They are isolated from the Martian weather by a Wind and Thermal Shield (WTS). A flexible cable connects the instruments to the E-box; a set of electronic cards located inside the lander thermal enclosure. Simultaneous measurements of pressure, temperature, wind and magnetic field will support the SEIS analyses.  To achieve the mission goals, the seismometers must meet the total noise requirements shown in \fig{Requirements}.  Such noise levels have been obtained on Earth with similar wind shielded surface seismometers \citep{lognonne96}. For more details about the SEIS instrument see \cite{lognonne2015b}.

 \figuremacroW{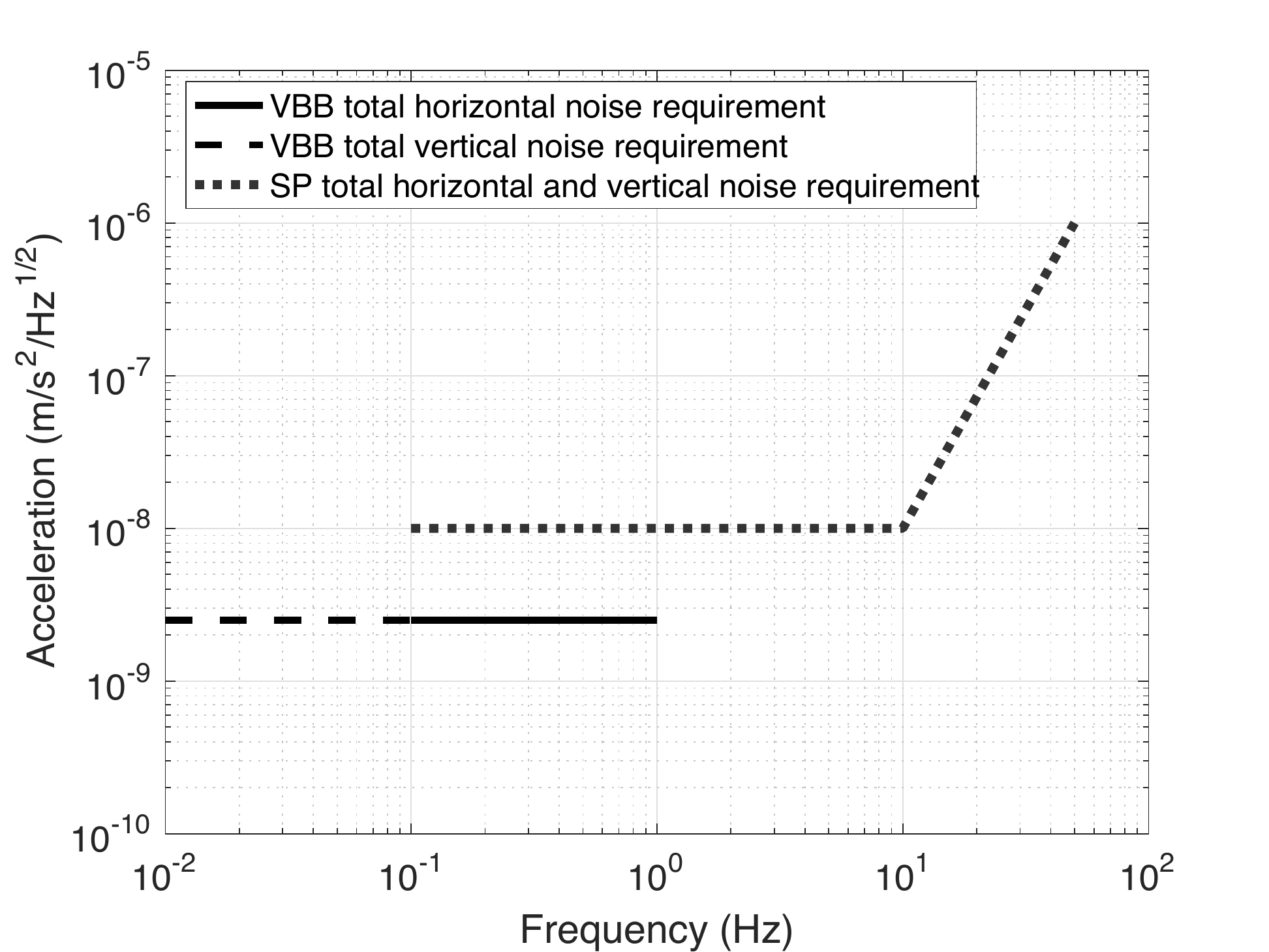}{SEIS noise requirements}{The total noise requirements for the very broad band (VBB) seismometer and for the short-period (SP) seismometer on Mars are shown by the black lines (solid for the horizontal requirement, dashed for the vertical requirement) and grey dotted lines, respectively. }{0.8}

There are many potential sources of noise on seismic instruments. Some of these noise sources have been the study of detailed investigation \citep[the pressure noise, for example; see ][]{sorrells1971, sorrells1971b, lognonne93,murdoch2016b} and others, such as lander thermal crack noise produced due to the large diurnal temperature variations, have been observed directly during the Apollo program \citep[\eg][]{duennebier74}.  For an overview of all of the noise sources that may influence the SEIS instrument see \cite{mimoun2016}. Here we attempt to understand and evaluate the mechanical noise contributions on the SEIS instrument due to the interaction between the InSight lander and the Martian winds.  We also consider the wind mechanical noise from the other surface elements: the WTS, and the Heat Flow and Physical Properties Package \citep[HP3; the second InSight instrument]{Spohn2014}.

Wind induced noise has been directly detected by the Viking seismic experiment \citep{anderson1977, nakamura1979}.  In fact, significant periods of time during the Viking lander missions were dominated by the wind-induced lander vibration \citep{Goins1979}. The lander was indeed subject to lift forces generated by the wind and, consequently, its platform was moving due to the low rigidity shock absorbers of the lander feet \citep{lognonne93}. The big difference, however, between the Viking experiment and InSight, is that InSight will position the seismometers directly on the Martian surface, rather than keeping them onboard the vibrating lander.   This lander mechanical noise has been recognised to be a potential problem for future space missions involving planetary seismometers, even when they are set on the ground \citep{lorenz2012}.  Like for Viking, the wind is expected to exert drag and lift forces onto the InSight lander and these stresses will be transmitted to the ground though the three lander feet, and then propagated through the ground to SEIS as an acceleration noise.  The same will occur for the WTS and the HP3 and efforts have recently been made to design a torque-less wind shield that could be used on Mars to reduce this effect \citep{nishikawa2014}.

In this paper we explain how we model the seismic noise that will be produced on SEIS as a result of the InSight lander vibrations. First, we analyse in-situ wind measurements from previous Mars space missions to understand the wind environment that we are likely to encounter. Next, we discuss the regolith properties on Mars before entering into a detailed discussion about the lander aerodynamics and our method to model how the stresses exerted on the ground at the lander feet are transmitted to, and registered on SEIS, as a seismic signal. We then present our noise maps that will be used to select the best deployment site for SEIS once we arrive on Mars. We then perform a Monte Carlo analysis to determine the sensitivity of our model results to the key uncertain parameters in the model, namely the environment variables.  Although the detailed simulation of the solar panel resonances are not included in our model, we analyse images from the Phoenix lander to predict typical solar panel resonant frequencies. Using the estimated resonant frequency, we demonstrate the influence that the InSight solar panel resonances will have on the SEIS seismic signal. Finally, we apply our mechanical noise model to estimate the noise produced on SEIS by the wind and thermal shield, and HP3 vibrations. 

As the very broad band seismometer is the critical instrument for achieving the Insight mission objectives \citep{banerdt2016}, we will concentrate on the [0.01-1 Hz] bandwidth. Additionally, our analyses will mostly be performed in the frequency domain as the SEIS noise-related requirements are defined as a function of frequency rather than time.

\section{Wind and dynamic pressure on Mars}

\subsection{Wind measurements on Mars}

The Phoenix Mars Lander was the first spacecraft to successfully land in a polar region of Mars (68.22$^\circ$ N, 125.75$^\circ$ W; \fig{MapMars}). The mission lasted 152 sols corresponding to $L_s$ = 76$^\circ$ to 148$^\circ$ . Wind speeds and directions at a nominal height of 2 m above the Martian surface were measured by a mechanical anemometer, the so-called Telltale wind indicator \citep[part of the Meteorological instrument packages; ][]{gunnlaugsson08, HolsteinRathlou10}. We use all of the Phoenix Telltale experiment data that are available on the Planetary Data System. 

The Viking 1 Lander touched down in western Chryse Planitia (22.70$^\circ$ N, 48.22$^\circ$ W; \fig{MapMars}). The Viking 2 Lander touched down about 200 km west of the crater Mie in Utopia Planitia (48.27$^\circ$ N, 225.99$^\circ$ W; \fig{MapMars}). On Viking Landers 1 and 2 a meteorology boom, holding wind direction, and wind velocity sensors extended out and up from the top of one of the lander legs (part of the Viking Meteorology Instrument System). The wind speed was measured by a hot-film sensor array, while direction was obtained with a quadrant sensor \citep{chamberlain76}. The Viking Lander 1 and 2 wind measurements were taken at a nominal height of 1.61 m \citep{tillman94}. The highest temporal resolution wind data available from the Viking Landers\footnote{These data were provided by J. Murphy and J. Tillman, via D. Banfield.} are available from Sols 1-49 for Viking Lander 1 (VL1) and Sols 1-127 for Viking Lander 2 (VL2). Given our interest in the mean wind properties, spurious data points (those exceeding several tens of ms$^{-1}$  for very short periods of time) are replaced by the mean wind speed.

\figuremacroW{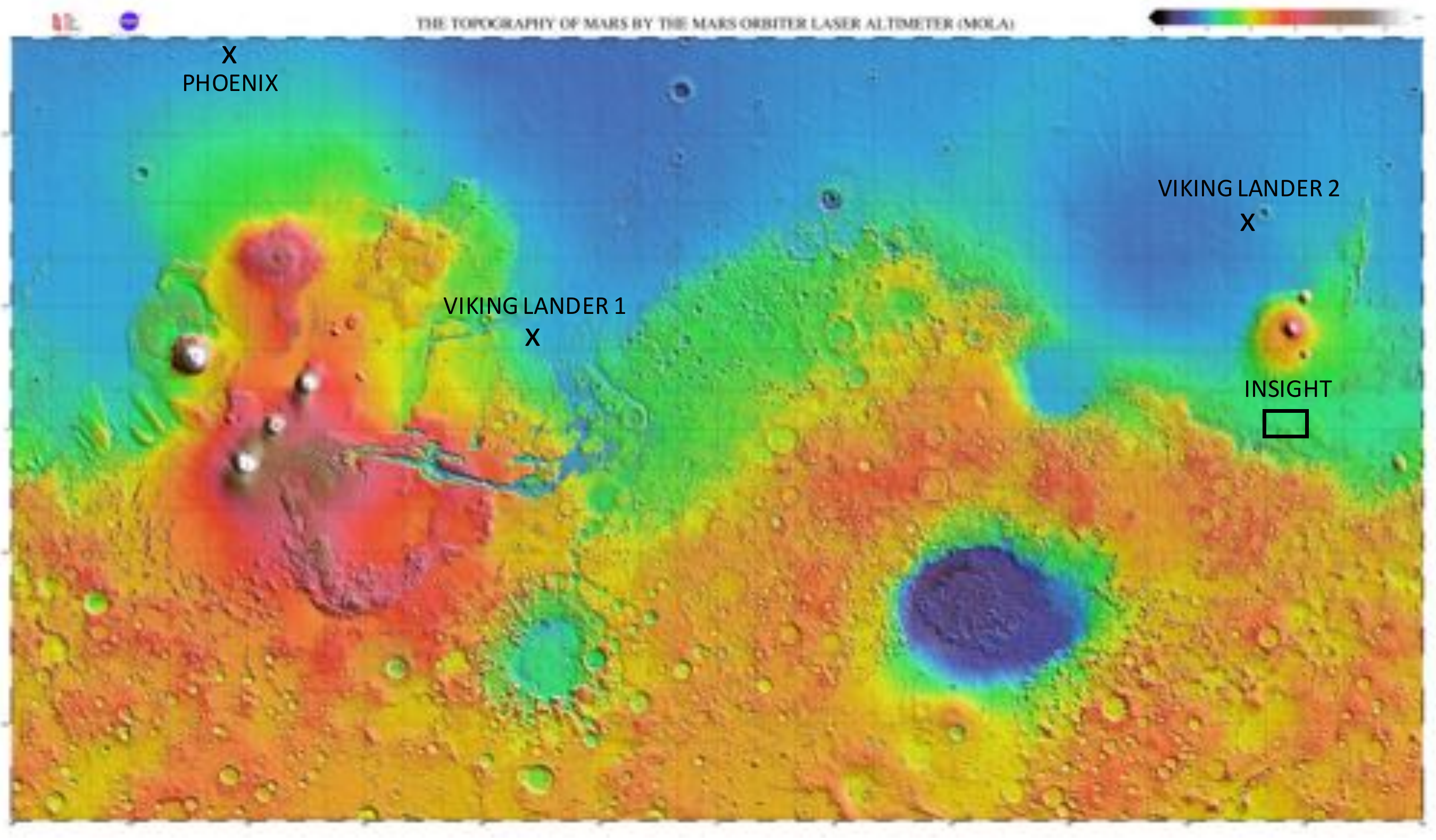}{Topography map of Mars showing lander locations}{The topography of Mars by the Mars Orbiter Laser Altimeter (MOLA). The colour code indicates the topography with dark purple being -9 km and white being +14 km.  The approximate locations of Viking Lander 1, Viking Lander 2 and Phoenix Mars Lander are shown. These are the three Mars landers discussed in detail in this paper. Also indicated is the InSight landing region. Map image credit: MOLA Science Team, MSS, JPL, NASA. }{0.8}

\subsection{Wind variation with height}\label{s:height}

The wind profile over an aerodynamically rough surface is given by \citep{prandtl1935, bagnold1941}:

\begin{equation}
U (z) = \frac{u^*} {\kappa} \ln(z/z_0)
\end{equation}

\noindent where $z$ is the vertical distance from the surface, $u^*$ is the wind shear velocity or friction velocity and is a measure of the gradient or fluid flow, $\kappa$ is the von Karman constant and is equal to 0.40, and $z_0$ is the surface roughness length. \Fig{HeightVariationNEW} gives an example of the wind variation with height for two different surface roughness lengths, assuming that the wind velocity at 1.6 m is 3.5 ms$^{-1}$  (the mean Martian wind speed measured by Phoenix and the two Viking Landers; \fig{WindSpeedLocalHourwMean}). Then, in order to scale the wind speeds measured at a given height to the height of the InSight lander, the Wind and Thermal Shield, or the HP3, we can express the wind speed at height $z$ as a function of the wind speed at the reference height $z_r$ and the surface roughness length $z_0$, i.e.,
	
\begin{equation}
U (z) = U (z_r) \frac{\ln(z/z_0) }{ \ln(z_r/z_0)}
\end{equation}

\noindent \cite{Sullivan00} determined the Martian surface roughness length to be $\sim$3 cm using the Imager for Mars Pathfinder (IMP). However, they note that the Pathfinder landing site is rockier and rougher than many other regions of Mars. Indeed, the InSight landing site is expected to be smoother, and in the following calculations the surface roughness length is assumed to be 1 cm based on the InSight Environment Requirements Document \citep{InSightERD}.

\figuremacroW{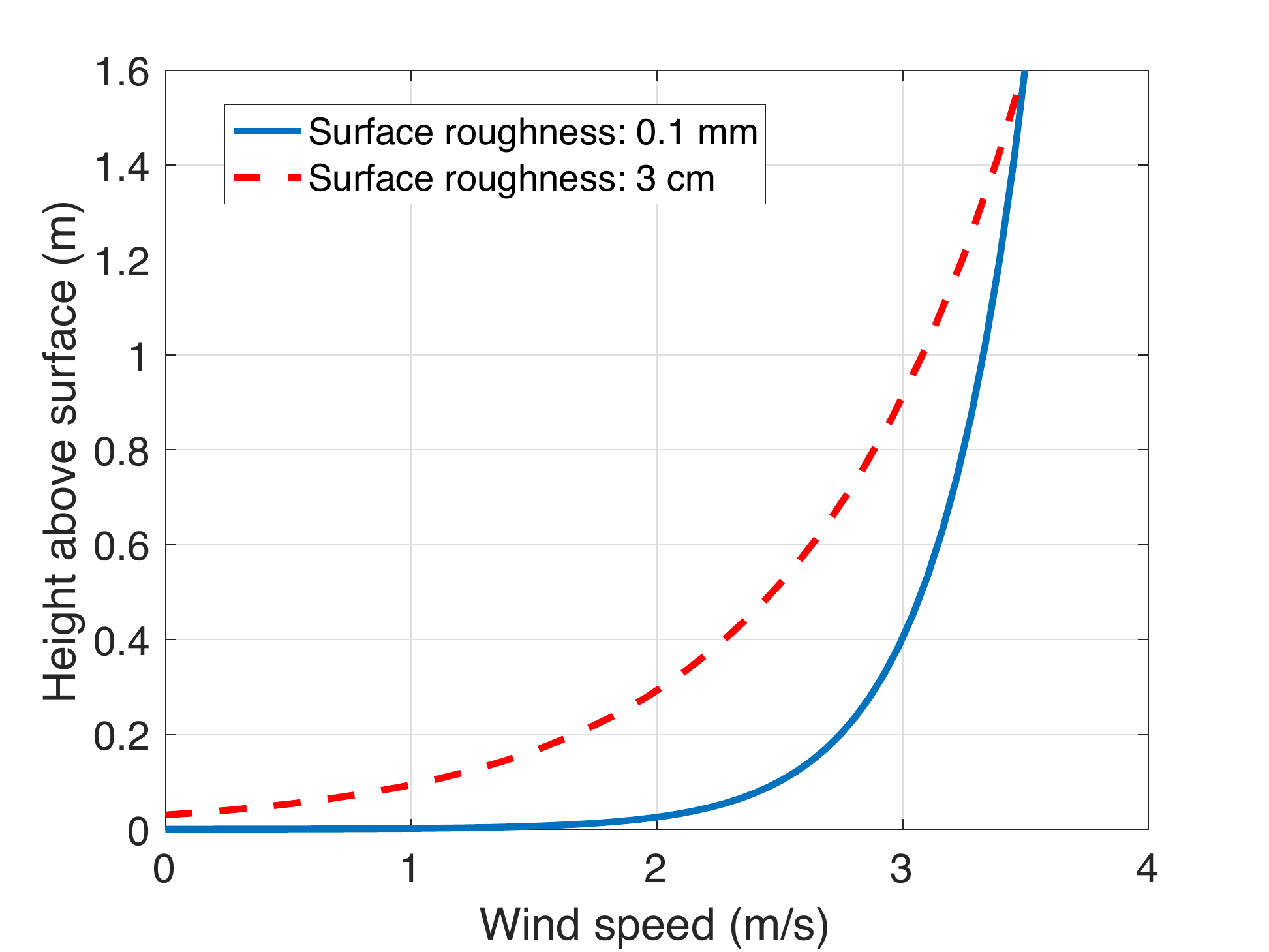}{Wind variation with height}{The wind velocity profile between the surface and the height at which the Viking Lander wind measurements were made. For this figure it is assumed that the wind velocity at a height of 1.6 m is 3.5 ms$^{-1}$ . The solid blue curve shows the wind profile for a surface roughness of 0.1 mm and the dashed red curve shows the wind profile for a surface roughness of 3 cm.}{0.8}

\subsection{Dynamic pressure spectral density}

The dynamic pressure ($P$) is calculated from the horizontal wind speed ($U$) and the air density ($\rho$) as follows:

\begin{equation}
P = \frac{ \rho {U}^2}{2}
\end{equation}

Then, using the equation above relating the wind speed at the reference height $z_r$ to the wind speed at height $z$, we can express the dynamic pressure at height $z$ as a function of the wind speed at the reference height and the surface roughness:

\begin{align}
&P(z) = \frac{\rho {U (z)}^2}{2}  \\
&P(z) = \frac{ \rho}{2}  \left [{\frac{\ln(z/z_0) }{ \ln(z_r/z_0)}} \right ]^2 {U(z_r)}^2 \\
&P(z) = C  {U(z_r)}^2 \\
\end{align}
where
\begin{align}
&C(z) =  \frac{ \rho}{2} \left  [{\frac{\ln(z/z_0) }{ \ln(z_r/z_0)}} \right ]^2  \\
\end{align}

The wind dynamic pressure, and thus the wind force, is directly proportional to ${U}^2$, the `wind speed squared'.  We can, therefore, calculate the wind speed squared amplitude spectral density (ASD) at the reference height of 1.61 m, and then scale this by $C$ to determine the dynamic pressure ASD at the required height. 

\subsection{Wind sensor instrument noise and available frequency range}\label{s:sensors} 

The sample interval varies throughout the data sets for each of the three space missions, the most common sampling intervals being 53 seconds for Phoenix, and from 2 to 64 seconds for VL1 and VL2. Therefore, to determine the wind speed squared ASD we first extract portions of data with approximately the same sampling interval. We then interpolate the wind speed data of each portion over a linear time array before squaring the wind speed data and performing the Fourier transform of the wind speed squared data. We acknowledge that, given the limited sampling frequencies of the existing data, there may be some non-linear effects that are not captured in the data.  

The Viking Lander wind speed measurement accuracy has been reported as $\pm$15\% for wind speeds over 2 ms$^{-1}$ \citep{chamberlain76, petrosyan11} and the Phoenix wind speed measurements were expected to be reliable in the 2-10 ms$^{-1}$ range \citep{gunnlaugsson08}. However, a simple description of the wind sensor's resolution and accuracy is not readily available \citep{HolsteinRathlou10, gunnlaugsson08,chamberlain76}. 

Therefore, to determine the highest frequency measurements that can be trusted in our data sets, we calculate the average spectrum of the ten longest continuous sequences in the combined data set.  We find that the spectrum levels out at frequencies higher than 0.02 Hz and this is probably indicating that the noise limit of the instrument sensitivity has been reached, or that the intrinsic response time of the instrument is longer than the shortest sampling interval. In consequence, during the subsequent analyses only data of frequencies up to 0.02 Hz will be considered. 

\subsection{Day/Night wind speed variations}

The wind speed on Mars varies as a function of local time (\fig{WindSpeedLocalHourwMean}). To determine the difference in the day and night time wind speed spectra we consider the local time of each of the wind speed time series extracted from the data and used to make the spectrum. Using the mean wind speed as a function of local hour (\fig{WindSpeedLocalHourwMean}), we define the day time as 6h-18h and the night time as 18h-6h. If a time series is entirely within the defined day time hours, the spectrum is defined as a daytime spectrum, and if a time series is entirely within the defined night time hours or covers both night and day time hours, the spectrum defined as a night time spectrum.  

\figuremacroM{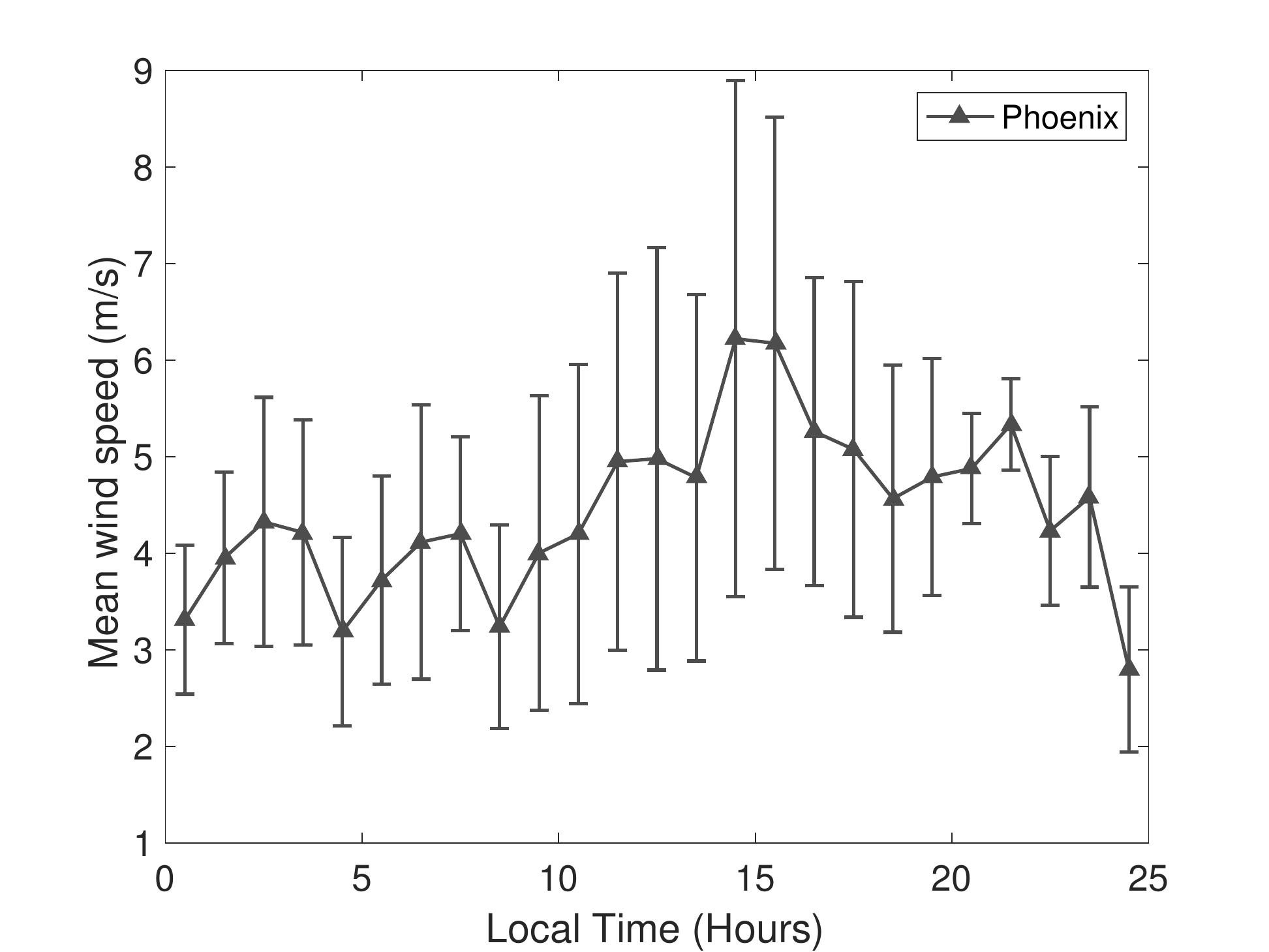}{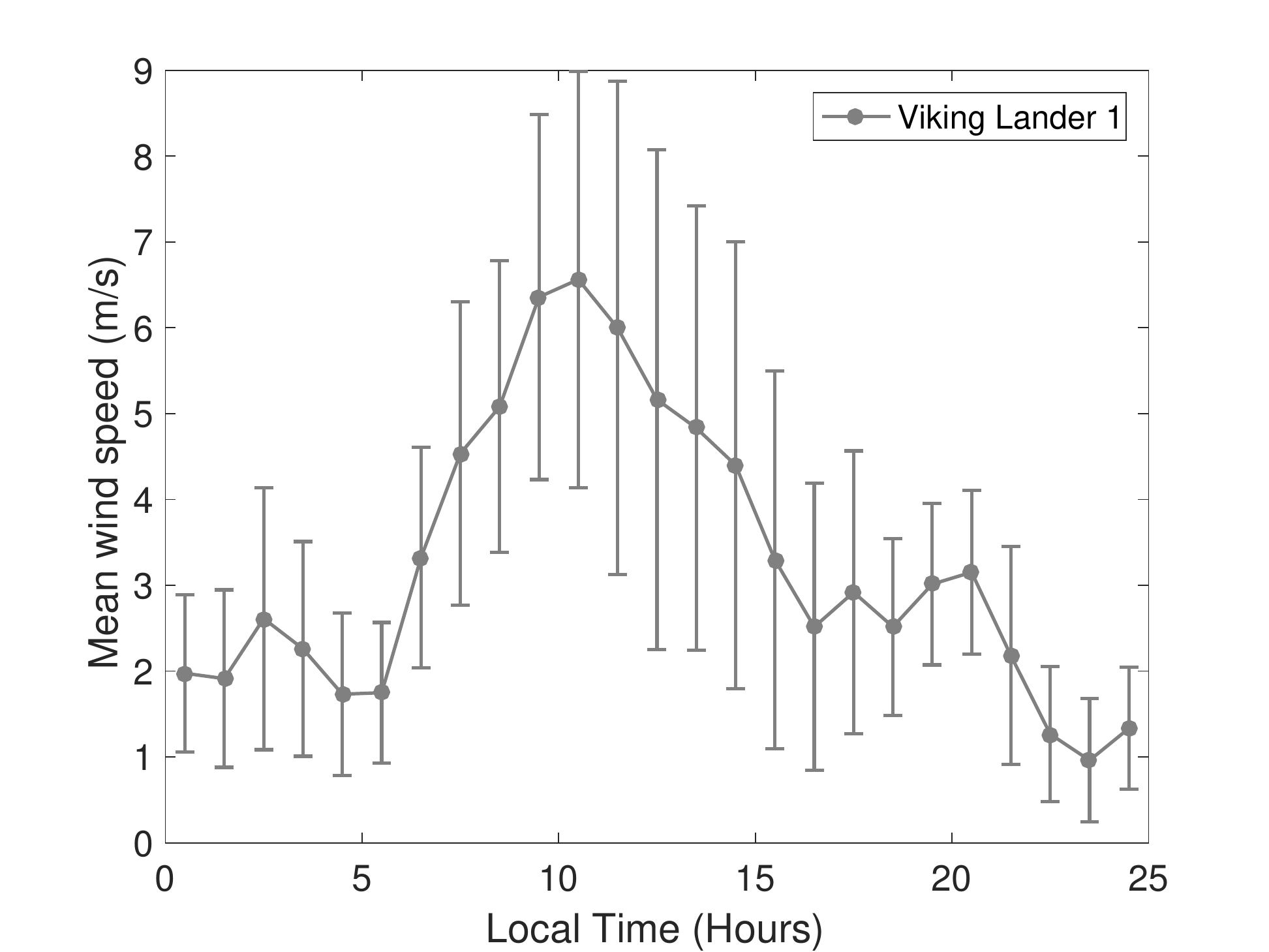}{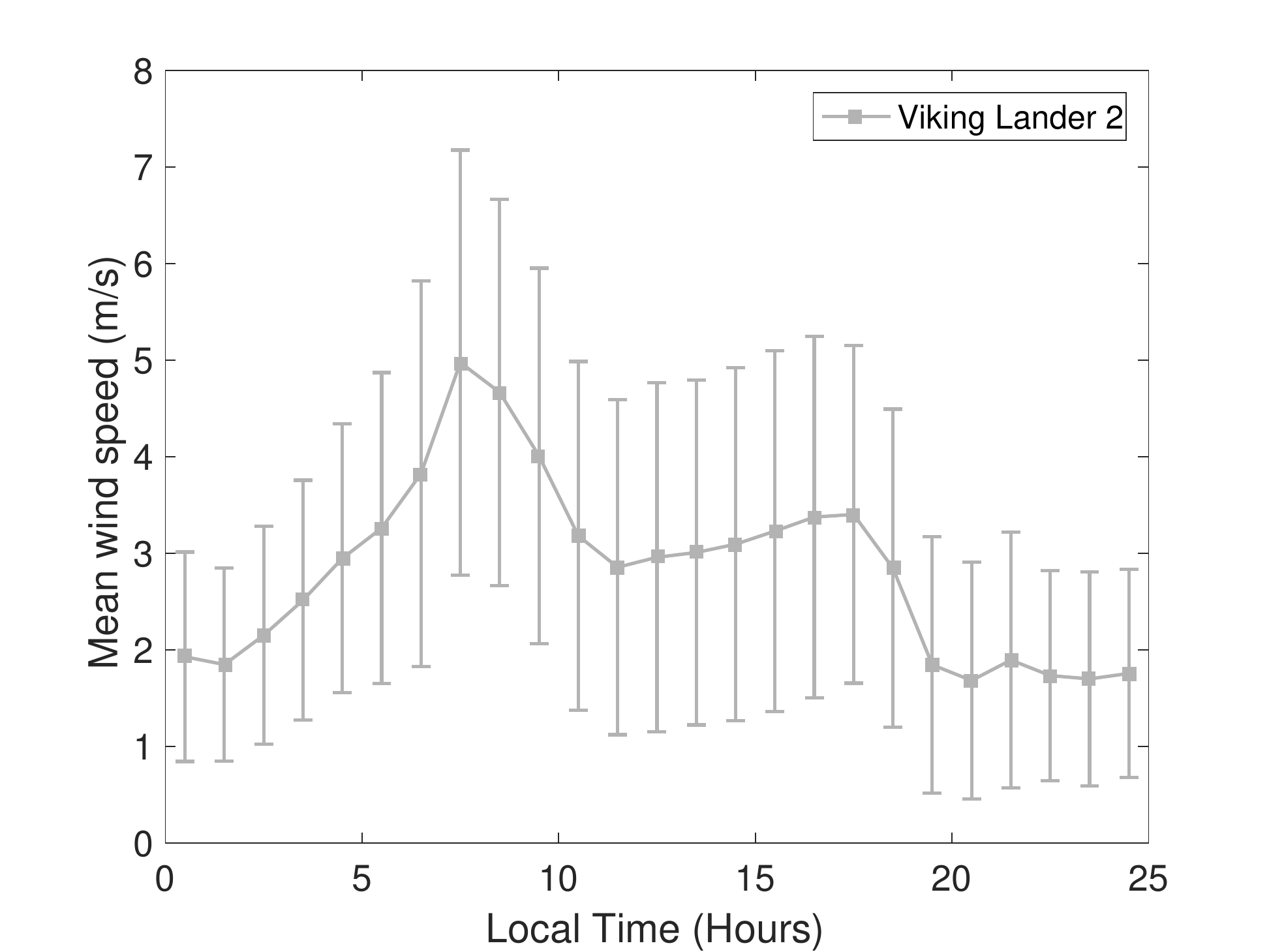}{WindSpeedLocalHourwMean}{Wind speed vs. local time}{Plot of the mean wind speed as a function of local time for the three data sets. The data are binned into 25 equal bins (equal duration) per sol. The error bars represent the standard deviation of the mean wind speed for each bin. The Phoenix wind data have been scaled to a height of 1.61 m to allow a direct comparison with the Viking Lander data (assuming $z_0$ = 1 cm). }

\subsection{Linear extrapolation for wind speed squared spectra at high frequencies}

To estimate the mechanical noise in the SEIS bandwidth it is necessary to perform a linear extrapolation of the wind speed squared ASD to higher frequencies (up to 1 Hz for the VBB bandwidth and 50 Hz for the SP bandwidth). However, determining an accurate high frequency spectrum to represent the wind on Mars is not easy.  The spectrum will change dramatically with time of day (due to atmospheric static stability and depth of the convective boundary layer) and wind speed (shifting the frequencies up and down for a given scale eddy).  In addition, the atmospheric circulation on Mars will be strongly impacted by atmospheric dust at all scales \citep{bertrand2014}.  One example of these daily variations - the turbulent wind behaviour - is demonstrated in \fig{Turbulence}.  For the Phoenix data set we observe higher levels of turbulence around midday, consistent with the results of \cite{HolsteinRathlou10}. The Viking Landers' results, however, seem to show more complex variations in turbulent behaviour. For both VL1 and VL2, the data appear to show three peaks in turbulence with two ``quiet'' periods in between: a peak is observed around midday (as for the Phoenix data) but two more peaks are observed in the early morning and in the late evening. However, it may be possible that the additional peaks in the Viking data are due to the reduced wind sensor performance at $<$ 2 ms$^{-1}$ (see \sect{sensors}) and not necessarily increased turbulence.
 
 \figuremacroM{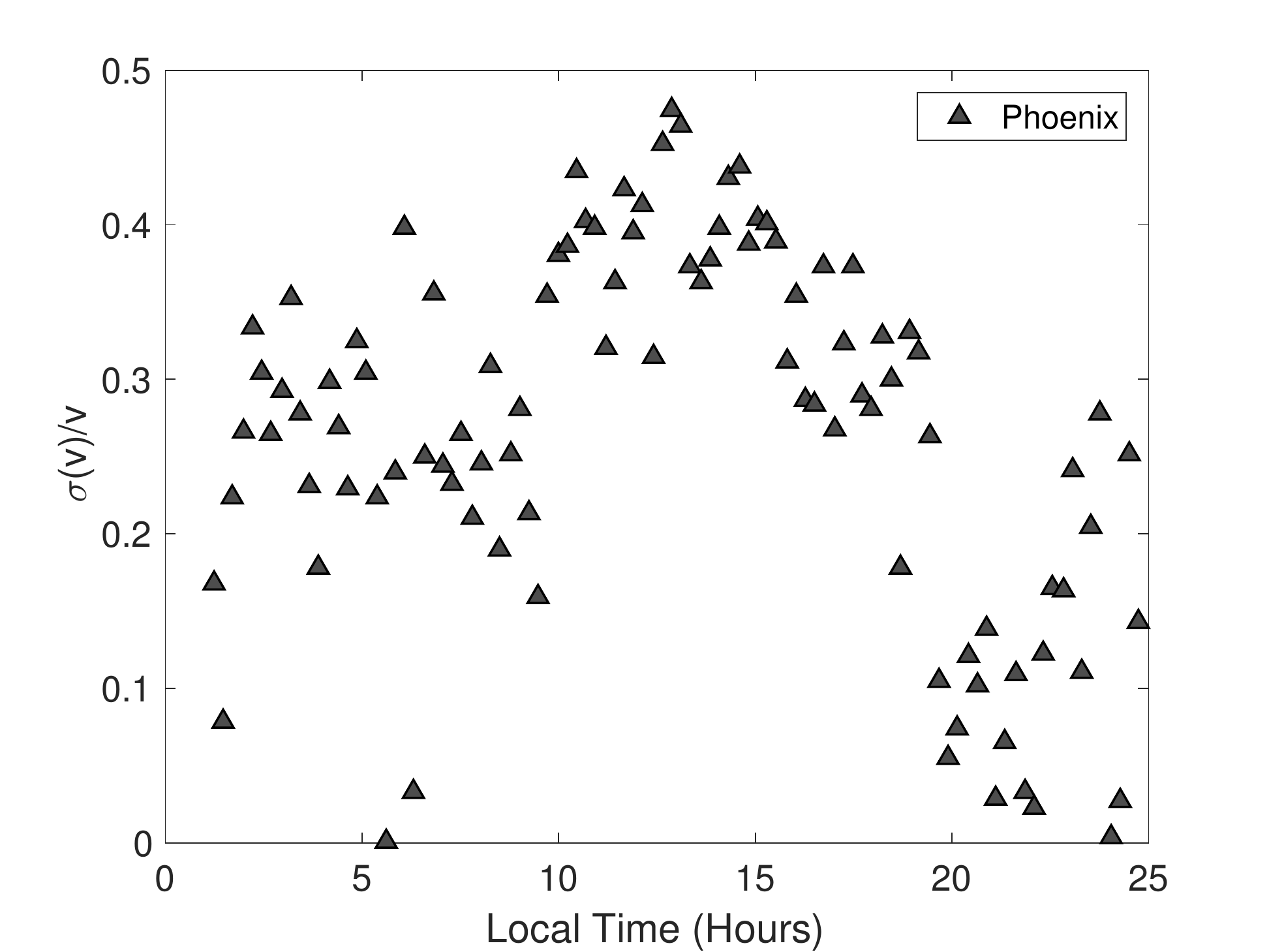}{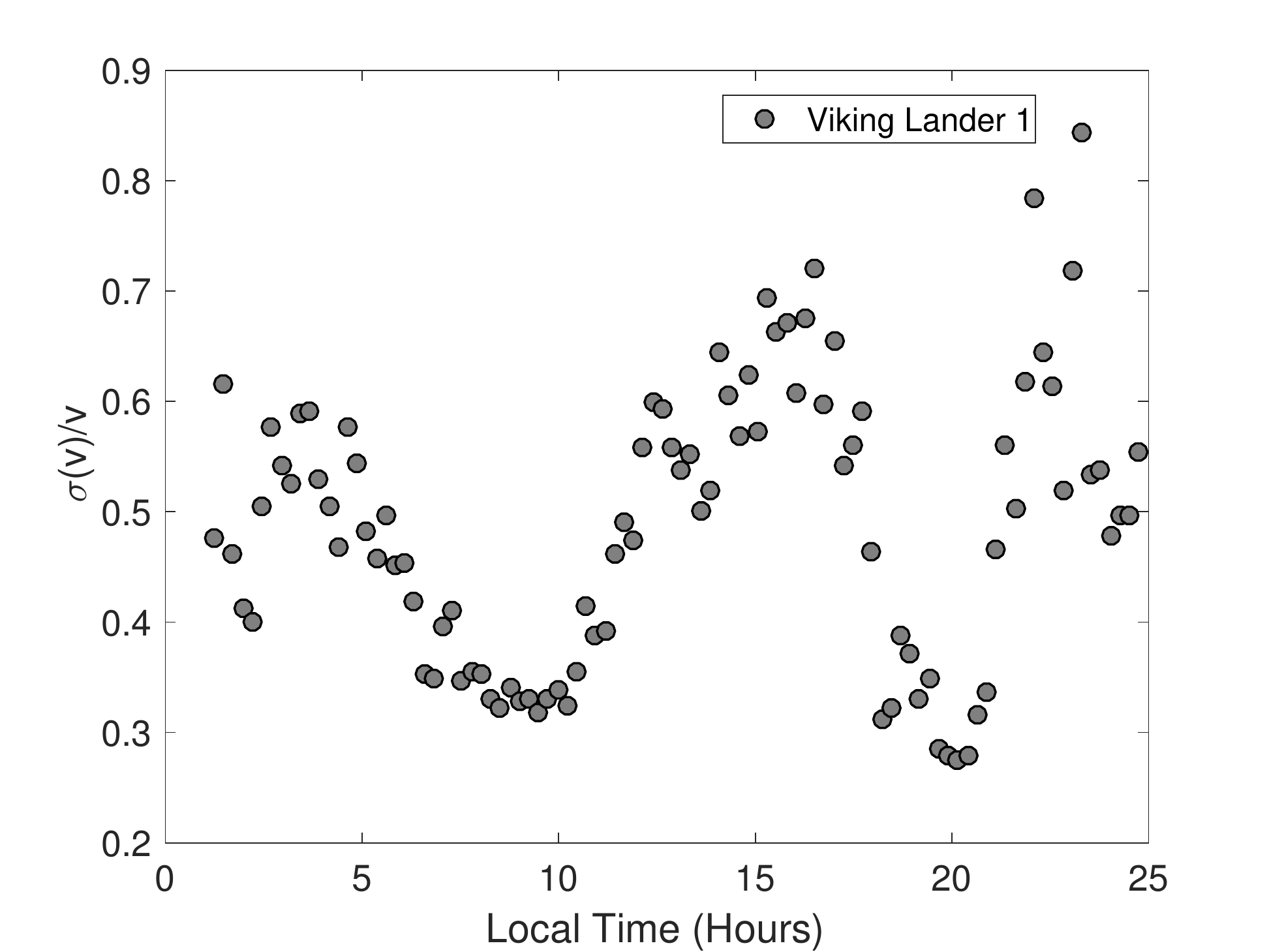}{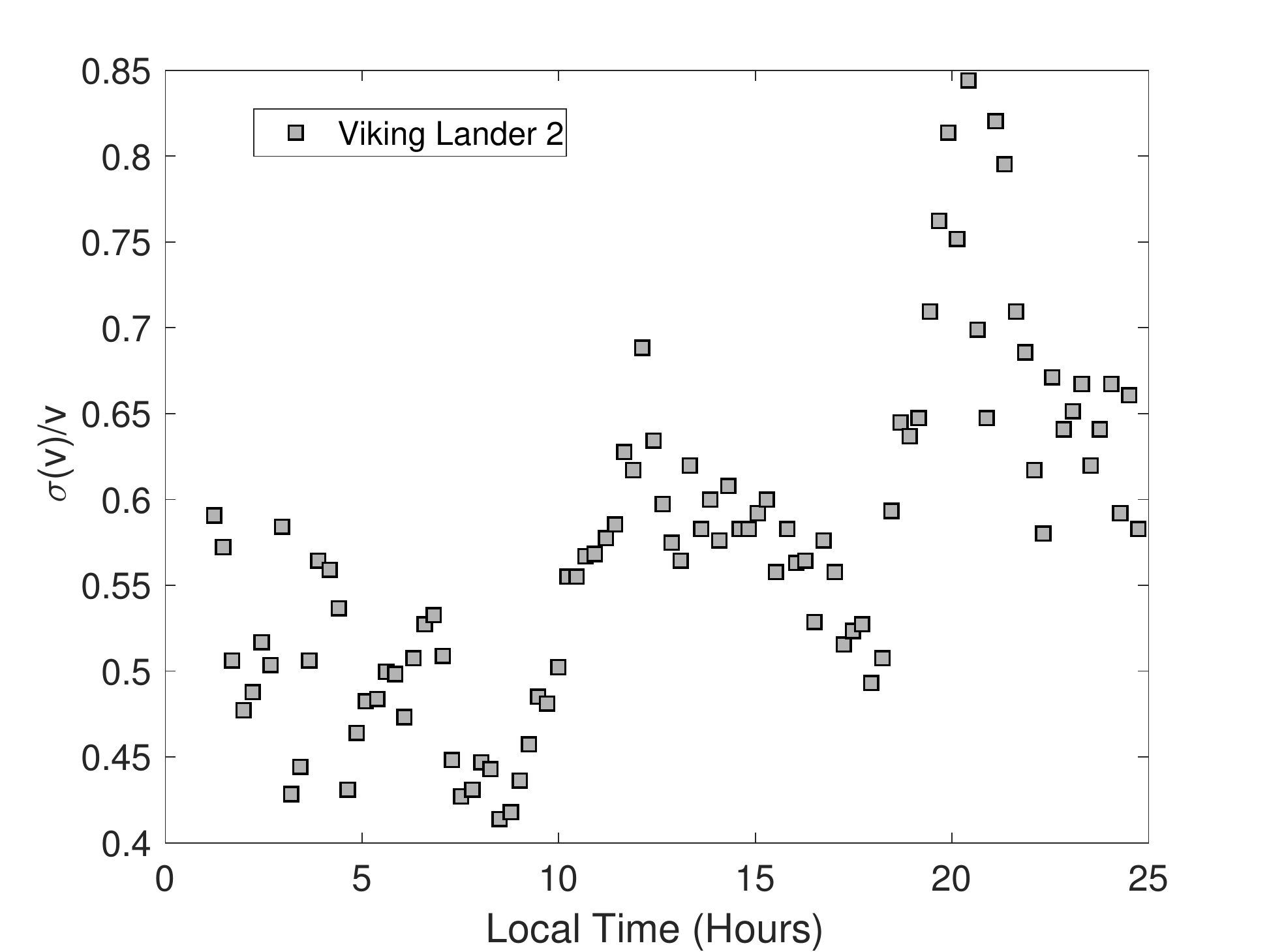}{Turbulence}{Turbulence vs. local hour}{Standard deviation of wind speeds divided by the average wind speed as a function of the local time for the Phoenix data (triangles), Viking Lander 1 data (circles) and Viking Lander 2 data (squares). This provides an indication of the periods of turbulent wind behaviour with the higher values indicating more turbulent behaviour of the wind \citep{HolsteinRathlou10}.}

The wind speed spectrum, therefore, contains many complexities and, currently, there are no in-situ measurements at frequencies in the bandwidth of SEIS.  Therefore, in the absence of sufficient in-situ data, we turn to theoretical arguments to propose a representative spectrum. 

The `source region' of turbulence, created by \eg solar radiation and atmospheric instabilities at large scales, is present at low frequencies. Most of the atmospheric kinetic energy is contained in these large-scale and slowly-evolving structures, and the spectrum should be relatively flat.  At intermediate frequencies energy cascades from these large-scale structures to smaller and smaller scale structures by an inertial mechanism (this is the `inertial regime'). To predict the spectral slope in the inertial regime we can use Kolmogorov's law, which states that the energy spectrum in the inertial regime has the form $E(k) = \alpha_1 \epsilon^{2/3}k^{-5/3}$ where $k$ is the wave number of the motion, $\epsilon$ is the turbulence energy dissipation rate and $\alpha_1$ is a dimensionless constant known as the Kolmogorov constant \citep{kolmogorov1941}. The spectral slope in the inertial regime is thus expected to be -5/3. 

The low frequency end of the inertial regime is set by the wind speed and the measurement height, which determines the typical dominant eddy size. As this is related to the boundary layer depth, it may also change as a function of the time of day. Given the lack of high frequency in-situ wind data (\sect{sensors}), we currently have no direct knowledge about where the transition from the source to the inertial regime occurs on Mars.  However, from their analyses of turbulence characteristics using Mars Pathfinder temperature fluctuation data, \cite{Schofield97} suggest that this transition occurs in the range of 10 and 100 mHz. A similar analysis using Phoenix data \citep{Davy2010} estimates this transition to be at $\sim$10 mHz. Based on this information, and the results of Large Eddy Simulations of candidate InSight landing sites \citep{kenda2016,murdoch2016b}, we place this transition at $\sim$15 mHz. 

The size of the smallest structures is determined when the inertial forces of an eddy are approximately equal to the viscous forces. This corresponds to when the turbulent structures are so small that molecular diffusion starts to become important. This is known as the Kolmogorov length. Eventually (at higher frequencies), the inertial regime moves into the dissipation regime and the spectrum should fall off very steeply because the viscosity strongly damps out the eddies.  The high frequency end of the inertial regime is, therefore, set by the Kolmogorov length which, due to the very low atmospheric density, is much larger on Mars than on Earth \citep{larsen02,petrosyan11}.  In consequence, the extent of the inertial regime is greatly reduced on Mars. It has even been suggested that the inertial regime is `virtually absent from the turbulence in the Martian atmospheric surface boundary at this height' \citep{tillman94}.  However, as the spectrum is likely to fall off with a slope steeper than -5/3 in the dissipation regime, we assume a worst case in which the inertial regime is present to the highest frequencies considered. 

Putting these arguments together allows us to suggest the following form for the wind speed squared spectrum linear model (defined at $z_r = 1.61$ m) as a function of frequency ($f$), amplitude ($B$), and with a cut-off frequency ($f_{cut}$) at 15 mHz. 

\begin{align}
& f < f_{cut}: {U}^2(f) = B \: {\textnormal m^2 \textnormal s^{-2} \textnormal Hz^{-1/2}}  \\
& f \geq f_{cut}: {U}^2(f) = B (\frac{f}{f_{cut}})^{5/3} \: {\textnormal m^2 \textnormal s^{-2} \textnormal Hz^{-1/2}}
\end{align}

To determine the amplitude of the linear wind speed squared model we consider the distribution of the day and night wind speed squared amplitudes in the [1 - 15 mHz] bandwidth. The day and night data can be approximated by lognormal distributions (\fig{PDF}). The cumulative probability distribution of amplitudes can then be calculated allowing an estimation of the amplitude of the upper limits for the night and day spectra 50\%, 70\% and 95\% of the time. These resulting spectral amplitudes are shown in \tbl{Amplitudes}, and the complete linear models are shown in \fig{NewWindLinearModelFigwLES}. \cite{Lorenz1996} and \cite{Fenton2010} have previously suggested that the Weibull distribution is a flexible and accurate analytic description of wind distributions on Mars. However, for our particular data set, the lognormal distribution was a found to fit the data more accurately.

In reality the spectrum is likely to be more complicated than this simplistic model and we hope that future space missions (including InSight) will provide valuable data that will lead to a better understanding of the Martian atmosphere and allow these models to be improved. 

\figuremacroW{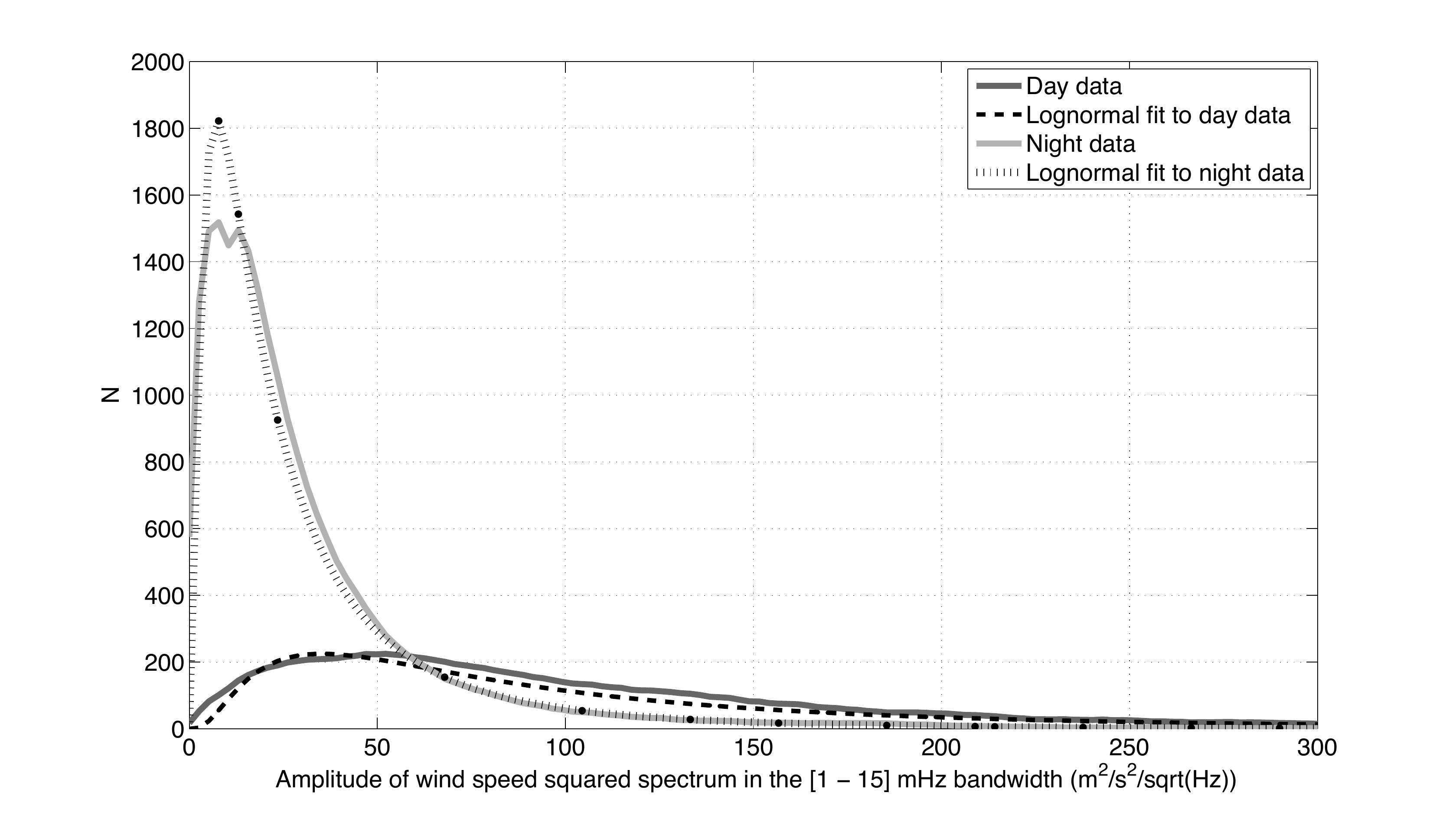}{Probability distribution function of the amplitudes of the wind speed squared spectra}{The probability distribution function of the amplitude of the day (dark grey) and night (light grey) wind speed squared spectra in the [1 - 15 mHz] bandwidth.  The dashed lines show the lognormal fits to the two datasets.}{0.8}

\begin{table}[htp!]\footnotesize
\caption{Linear wind speed squared model amplitudes at $f <$ 15 mHz (see \fig{NewWindLinearModelFigwLES})}
\begin{tabularx}{\textwidth}{ l   c  c  c  | c c c }
 & & \bf{Day}  & & & \bf{Night} &  \\
 & \bf{50\%}  & \bf{70\%}  & \bf{95\%} & \bf{50\%} & \bf{70\%}  & \bf{95\%} \\
 \hline
Amplitude, $B$ (m$^{2}$ s$^{-2}$ Hz$^{-1/2}$) & 78 & 125 & 345 & 21 & 34 & 105 \\
 \hline
\end{tabularx}
\label{t:Amplitudes}
\end{table}

\figuremacroW{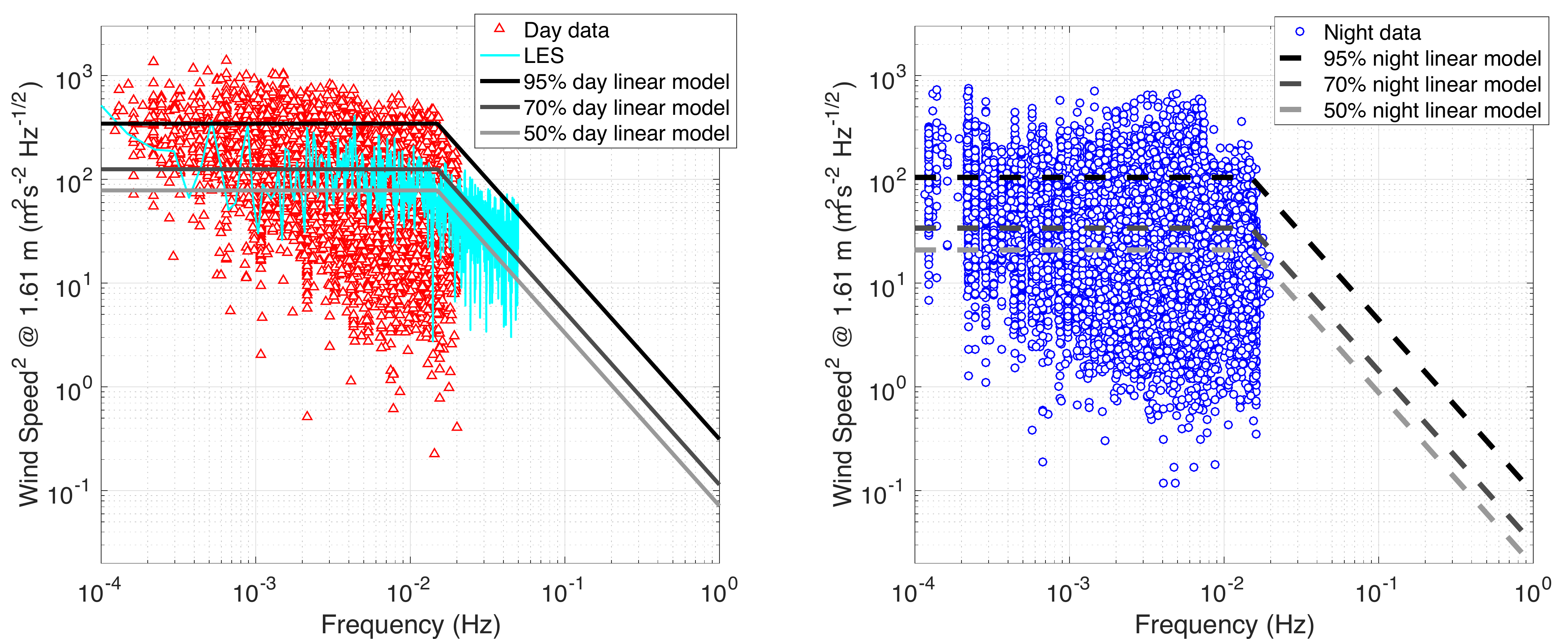}{Day/Night Wind Speed Squared Spectra Linear Models}{The wind speed squared spectra calculated from wind speed data from all of the data sets - Phoenix data, VL1 data and VL2 data. For clarity, only the points of the spectra are shown rather than the lines. If a time series is entirely within the defined day time hours the spectrum is shown in red triangles (left figure) and if a time series is entirely within the defined night time hours or covers both night and day time hours the spectrum is shown in blue circles (right figure).  The Phoenix wind data have been scaled to a height of 1.61 m to allow a direct comparison with the VL data (assuming $z_0$ = 1 cm). Also shown are the 50\% (light grey), 70\% (dark grey) and 95\% (black) linear models for the day (solid lines, left figure) and night (dashed lines, right figure). The amplitudes at $<$15 mHz are defined based on the distribution of wind speed squared amplitudes in the [1 - 15 mHz] bandwidth as described in the text and shown in \fig{PDF}.  However, the `95\% linear models' encompass $\sim$96\% of day time data points and $\sim$94\% of night time data points over the full frequency range. The day time wind speed squared spectrum for a candidate InSight landing site, calculated using Large Eddy Simulations, is shown in cyan in the left figure \citep{kenda2016,murdoch2016b}.}{1.0}

\subsection{Wind direction at the InSight landing site}

Elysium Planitia, the InSight landing site, is located in a place where different large scale wind currents meet \citep{bertrand2014}. The way in which they interact depends on the time of year, and the winds also exhibit a strong diurnal cycle of wind direction caused by thermal tides. The Mars Climate Database version 5.2 \citep{millour15} indicates an average large-scale wind from the North-West for the 2018 landing season (L$_s$=295$^\circ$).  We, therefore, assume a wind from the North-West (\ie towards the South-East) as the most common wind direction for this study. Then, later, we vary the wind direction through 360$^\circ$ as part of a Monte Carlo study considering the sensitivity to the environment parameters.  

The wind direction assumption is important for the decomposition of the drag force into the two horizontal components, and for the calculation of the deployment site noise maps (\sect{maps}).  In the baseline landed configuration, InSight will be aligned along the North-South axis with the deployment zone to the South (\fig{landerdeploy}). In consequence, when the wind comes from the South, SEIS is upwind of the lander and when the wind comes from the North, SEIS is downwind of the lander.  The azimuth of InSight after landing on Mars will be known and thus its position with respect to SEIS will be determined to within about $\pm$20$^\circ$. The correct azimuth can, therefore, be taken into account upon arrival to Mars.

\section{Martian regolith properties}\label{s:ground}

The seismic velocities of Martian regolith simulant (Mojave sand), measured in laboratory tests at a reference pressure ($p_{ref}$) of  25 kPa \citep[the smallest isotropic confinement stress used in the experiments;][] {delage2016}, are given in \tbl{regolith}. 

 \begin{center}
\begin{table}[h]
\caption{Insight landing site regolith properties valid for the upper 2 m of regolith measured at a reference pressure of 25 kPa. The regolith simulant used here is Mojave sand. The error is the standard deviation of the laboratory measurements. For details see \cite{delage2016}. }
\begin{tabular}{ccc}
 \hline
 \bf{Bulk density, $\rho_r$} & \bf{S-wave velocity, $v_S$} &  \bf{P-wave velocity, $v_P$} \\  
 \bf{(kg m$^{-3}$)} & \bf{(m s$^{-1}$)} & \bf{(m s$^{-1}$)} \\
 \hline
1665 $\pm$ 38 & 150 $\pm$ 17 & 265 $\pm$ 18 \\
\hline
\end{tabular}
\label{t:regolith}
\end{table}
\end{center}

However, on  the  surface  of  Mars  the  pressure  will  be  different   and   the   regolith   properties  will   change   accordingly.   The  pressure  under  each foot of the lander on  Mars ($p$) is   calculated   taking   into   account  the Martian surface gravity ($g$), the lander's mass ($m$),  the lander's foot  radius  ($r$) and  the  number  of  feet the lander has ($N$):

\begin{equation}
p = \frac{m g}{N \pi r^2}
\end{equation}

The $P$- and $S$- wave velocities directly under each foot on Mars are then calculated by extrapolating the ``reference'' $P$- and $S$- wave velocities ($v_{P_{ref}}$, $v_{S_{ref}}$) to the values at the required pressure, $p$. The extrapolation is performed assuming the following power law based on laboratory measurements \citep{delage2016}:

\begin{align}
v_P &= v_{P_{ref}} (\frac{p}{p_{ref}})^{0.3} \\
v_S &= v_{S_{ref}} (\frac{p}{p_{ref}})^{0.3}
\end{align}
  
The Young's modulus ($E$), shear modulus ($\mu$) and Poisson ratio ($\nu$) can then be calculated using the regolith bulk density ($\rho_r$) and the $P$- and $S$- wave velocities:

\begin{align}
E &= \rho_r v^2_S \frac{3 v^2_P - 4 v^2_S}{v^2_P - v^2_S} \\
\mu &= \rho_r v^2_S \\
\nu &=\frac{v^2_P - 2 v^2_S}{2(v^2_P - v^2_S)}
\end{align}

We note that the effective Young's modulus between the lander feet and the regolith is dominated by the Young's modulus of the regolith and depends weakly on the lander feet radius. 

\section{Calculating the lander mechanical noise}

The dynamic pressure of the wind will produce stresses on the lander body. These stresses will subsequently deform the ground resulting in a ground motion that will be registered on the seismometers. The magnitude of these stresses exerted on the ground by the lander feet depends on the dynamic pressure, the aerodynamic properties of the lander and the angle of attack of the wind. The proximity of SEIS to the lander noise source is such that no propagation effects are significant and that the noise is mostly static loading.

The baseline deployment configuration for SEIS, the wind and thermal shield (WTS) and the Heat Flow and Physical Properties Package \citep[HP3, the second InSight instrument;][]{Spohn2014} is given in \fig{landerdeploy}.

\figuremacroTwo{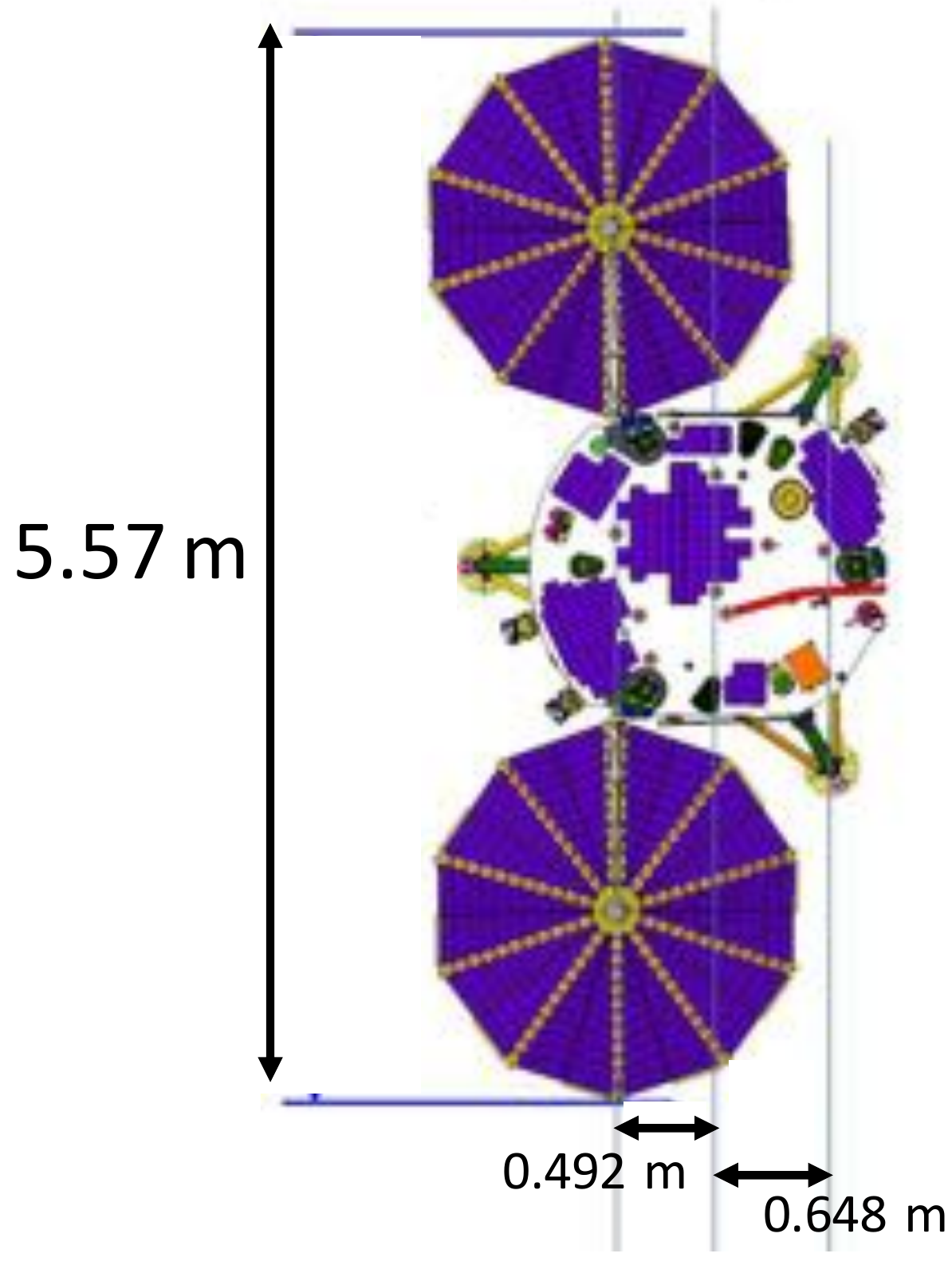}{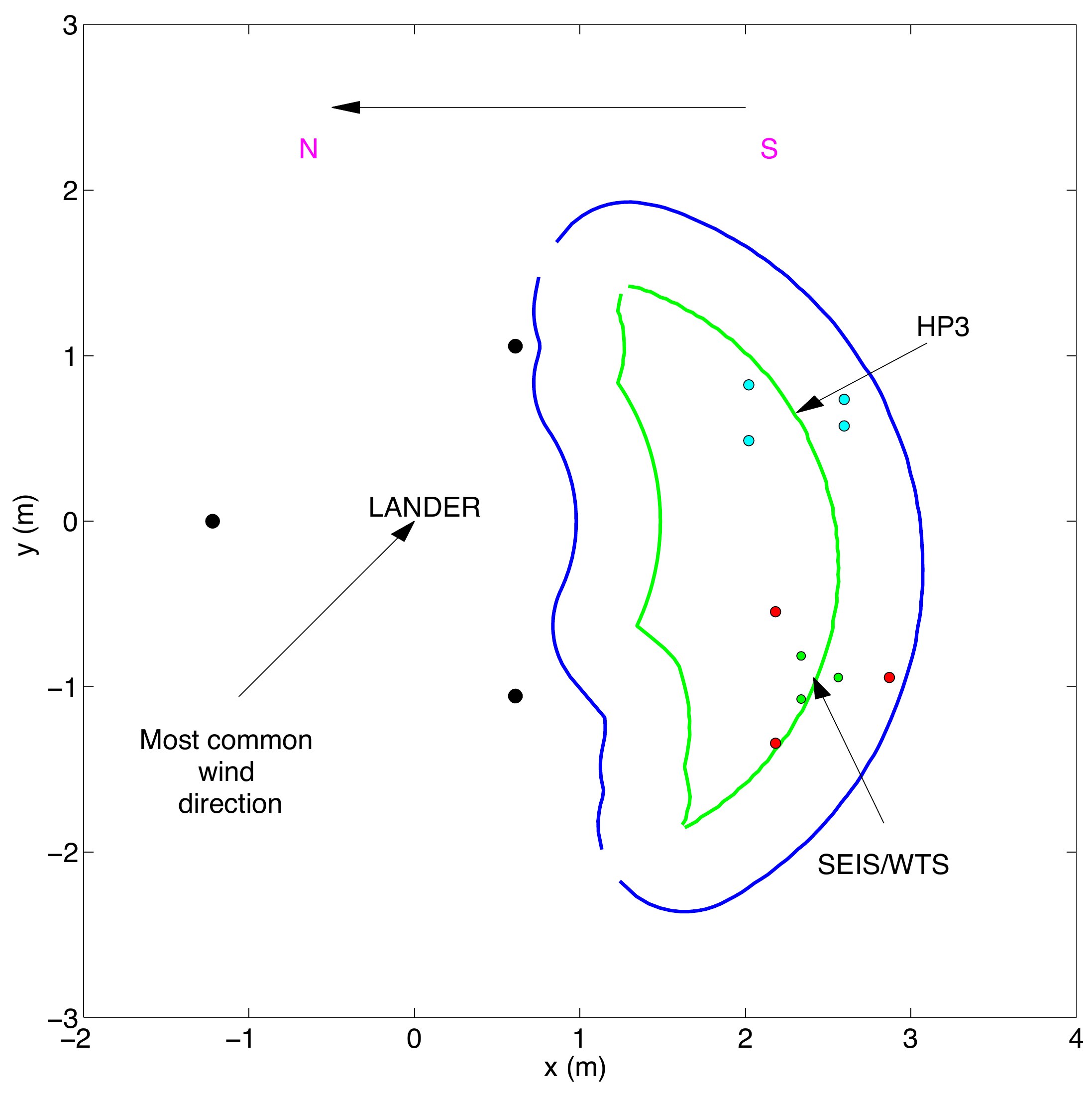}{landerdeploy}{The InSight lander and baseline deployment configuration}{(Left) The lander dimensions are provided in the diagram. The solar panels are offset with respect to lander body by 49.2 cm. Assuming that the geometric center of the lander at ground level is (0,0,0) the lander center of gravity (CoG) is at  (-0.038, 0.001, 0.777). (Right) The three black circles indicate the location of the three lander feet. The deployment zone is to the south of the lander. The blue and green lines shown the possible deployment zones for the WTS and SEIS, respectively. The baseline deployment locations of SEIS and the WTS are shown by the green and red circles, and the baseline deployment location of HP3 is shown by the cyan circles. The figure is to scale. The dominant wind direction is expected to be north-westerly. }

\subsection{Lander aerodynamics}

A diagram of the InSight lander is shown in \fig{landerdeploy}.  The main leg element of the InSight lander has a crushable honeycomb element in the load path that crushes based on the load encountered at impact. The two leg stabilisers are attached to stainless steel load limiter pins that bend on impact to further limit the load. This design is quite different from the Viking lander design, where the feet had spring-like structures. The InSight lander will have resonance modes but these are required to be at frequencies above 1 Hz. Consequently, we consider the lander, deck and legs, as an inelastic structure in the frequency band from 10$^{-3}$ to 10 Hz, and we do not include a detailed simulation of the resonances. We will, however, return to the question of the solar panel resonances in \sect{resonances}.

The lift and drag forces exerted on the lander are then given by:
 
 \begin{align}
F_l &= P  S_{Lift} C_l \\
F_d &= P S_{Drag} C_d \\
\end{align}

where $P$ is the dynamic pressure defined above, $S_{Lift}$ is the surface of the lander exposed to the lift force,  $S_{Drag}$ is the surface of the lander exposed to the drag force and $C_l$ and $C_d$ are the lander lift and drag coefficients, respectively. The lander lift and drag coefficients as a function of vertical angle of attack of the wind were provided by JPL following a series of wind-tunnel tests (\fig{aerodynamiccoeffs}). For the calculation of the coefficients, we assume the worst case vertical angle of attack of the wind for the InSight lander, which is expected to be 15$^\circ$. The lift force is assumed to act vertically only, and the drag force is decomposed into $x$ and $y$ components based on the wind direction.

This is a simplified approach to calculating the aerodynamic forces acting on the lander that does not take into account the wind interaction with the lander body \ie vortexes created by the lander body that could increase or decrease the lift and drag forces in certain configurations. 

\figuremacroW{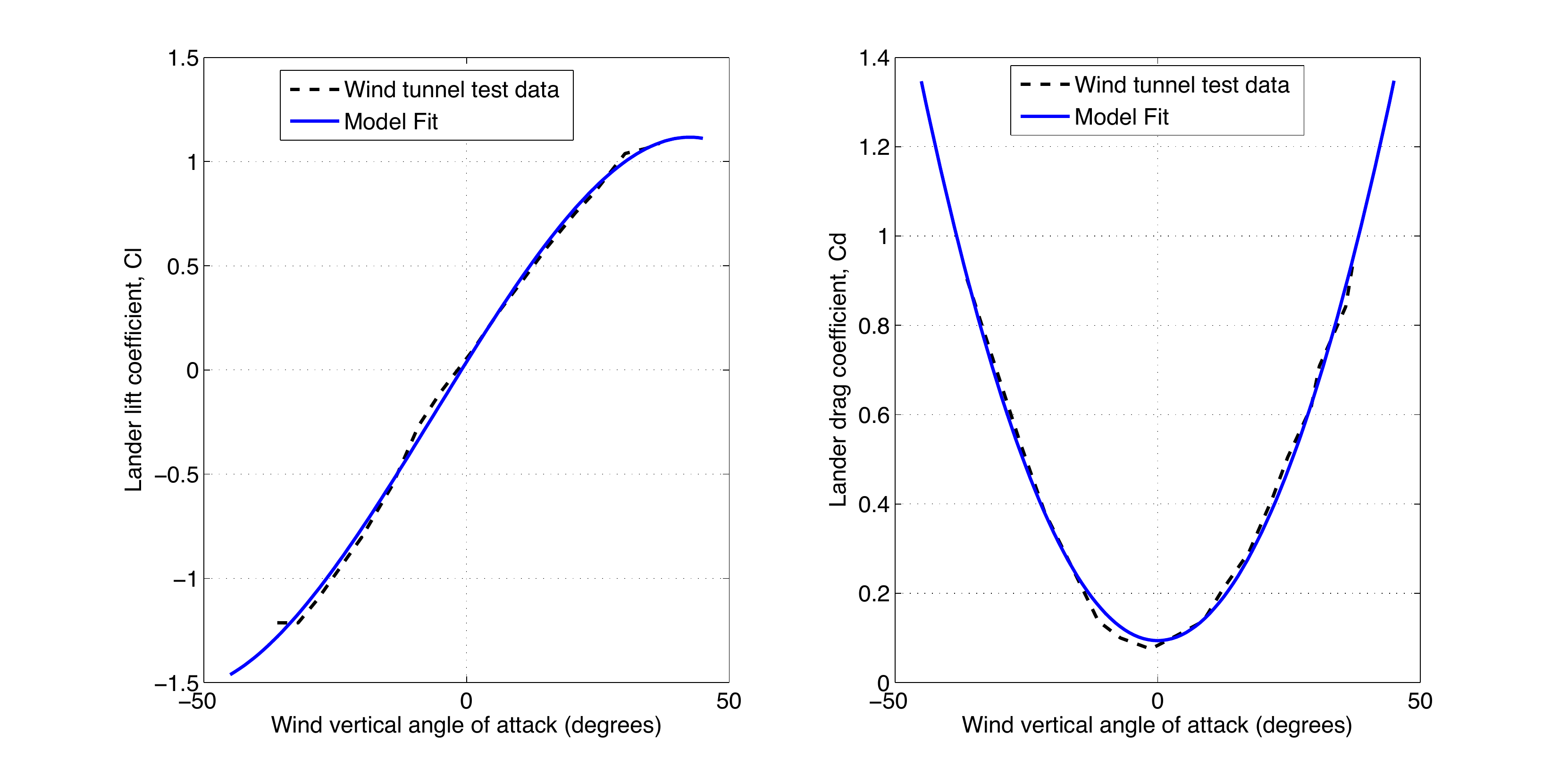}{Lander lift and drag coefficients as a function of vertical angle of attack}{Lander lift and drag coefficients as determine by wind tunnel tests are given by the dashed black line. The surface area of the lander exposed to the lift and drag forces is 7.53 m$^2$. In our model we use the best fit to this data, given by the solid blue line. The fits are described by the following analytic functions: $C_{l_{fit}} =  -5.7731 \times 10^{-6} \theta^3  -1.0581 \times 10^{-4}  \theta^2 + 0.0403  \theta +   0.0389$ and $C_{d_{fit}} =  6.1896 \times 10^{-4} \theta^2 +  1.1829 \times 10^{-5} \theta +   0.0938$, where $\theta$ is the vertical angle of attack.}{1}

The total variation in force exerted on the ground by the lander feet as a result of the aerodynamic forces is a combination of the effect of the lift and drag forces.  The distribution of loads at the three lander feet as a result of these forces is solved for by defining the positions of the three lander feet and the point of application of the aerodynamic forces (\ie the aerodynamic center of pressure) and assuming that the lander is in mechanical equilibrium. 

Assuming that the geometric center of the lander at ground level is (0,0,0) the lander center of gravity (CoG) is at  (-0.038, 0.001, 0.777). We assume that the height of the aerodynamic centre of pressure of the lander is at the height of the lander centre of gravity.  The aerodynamics of the solar panels will determine the horizontal coordinates of the lander centre of pressure. The center of pressure of each of the solar panels is half way between the edge and the center of the solar panels \ie at the 1/4 chord location ($\lambda$). This will cause a horizontal offset of the lander center of pressure in the direction of the incoming wind.  The solar panel offset ($x_o$) with respect to the geometric center of the lander body as described in \fig{landerdeploy} must also be taken into account. The coordinates of the aerodynamic centre of pressure of the lander ($CoP$) can be defined as follows:

 \begin{equation}
CoP = (-\cos(\beta)*(\lambda/4) + x_o, -\sin(\beta)*(\lambda/4), 0.777)
\end{equation}

However, when $\cos(\beta) = 0$ or $\sin(\beta) = 0$, we use the $x$ and $y$ coordinate of the centre of gravity of the lander, respectively. The horizontal wind direction $\beta$ is defined in an anti-clockwise direction with respect to the $x$-axis shown in \fig{landerdeploy} i.e., when $\beta$ = 0$^\circ$, the wind is in the +$x$ direction and when $\beta$ = 90$^\circ$, the wind is in the +$y$ direction.

\subsection{Ground deformation}

To understand the influence of the lander mechanical noise on the seismometers, we consider the deformation of the ground under the SEIS feet as a result of the stresses being applied at the lander feet.   Given the small distances between lander and SEIS feet compared to the thickness of the regolith layer it is possible to model the ground as an elastic half-space with properties of a Martian regolith.   We then use the Boussinesq point load solution \citep{boussinesq1885} to determine the deformation of the elastic medium caused by forces applied to its free surface.  Given the seismic velocities in \tbl{regolith}, typical wavelengths of seismic propagations are $\sim$30 km to $\sim$150 m in the [0.01 - 1] Hz bandwidth. As these distances are approximately 10 times or more larger than the typical distance between the lander feet and the SEIS feet, the static deformation hypothesis is a reasonable approximation. At higher frequencies, the lander-SEIS distance becomes comparable to the typical wavelengths of seismic propagations and the static deformation hypothesis may no longer be valid. 

Assume a point force ${\bf F}=F_1 {\bf e_1}+ F_2 {\bf e_2}+F_3 {\bf e_3}$ that is applied at the point ${\boldsymbol \xi} = \xi_1 {\bf e_1}+ \xi_2 {\bf e_2}+ \xi_3 {\bf e_3}$ and ${\bf \Lambda}= \Lambda_1 {\bf e_1}+ \Lambda_2 {\bf e_2}+ \Lambda_3 {\bf e_3}$  is some arbitrary point in the half-space $\Lambda_3 \ge 0$. The Green's tensor for displacements ($G_{ik}$), defined by the relation $u_i= \sum_{k} G_{ik}  F_k$, may be written in Cartesian coordinates as  \citep[solution from][]{landau1970}:

\begin{center}
\vspace{0.2 cm}
$G_{ik} = \frac{1}{4\pi \mu}$   
\begin{footnotesize}
$\left[\begin{array}{cccc}
\frac{b}{r} + \frac{x^2}{r^3} - \frac{ax^2}{r(r+z)^2} - \frac{az}{r(r+z)} & \frac{xy}{r^3} - \frac{ayx}{r(r+z)^2}  &  \frac{xz}{r^3} - \frac{ax}{r(r+z)}        \\
\frac{yx}{r^3} - \frac{ayx}{r(r+z)^2} & \frac{b}{r} + \frac{y^2}{r^3} - \frac{ay^2}{r(r+z)^2} - \frac{az}{r(r+z)}  &  \frac{yz}{r^3} - \frac{ay}{r(r+z)}         \\
\frac{zx}{r^3} - \frac{ax}{r(r+z)} & \frac{zy}{r^3} - \frac{ay}{r(r+z)}  & \frac{b}{r} + \frac{z^2}{r^3}
\end{array}\right]$
\end{footnotesize}
\vspace{0.2 cm}
\end{center}

where $x = \Lambda_1 - \xi_1$, $y = \Lambda_2 - \xi_2$, $z = \Lambda_3 - \xi_3$, and $r$ is the magnitude of the vector between ${\bf \Lambda}$ and ${\boldsymbol \xi}$, $a = (1-2\nu)$ and $b = 2(1-\nu)$, $\nu$ is PoissonÕs ratio and $\mu$ is the shear modulus (as defined in \sect{ground}).   For our calculations, we assume that $\Lambda_3 = 0$ and $\xi_3 = 0$ \ie the lander and SEIS feet are all on the surface of the regolith. The Green's tensor then simplifies to:

\begin{center}
\vspace{0.2 cm}
$G_{ik} = \frac{1}{4\pi \mu}$   
\begin{footnotesize}
$\left[\begin{array}{cccc}
\frac{b}{r} + \frac{x^2}{r^3} - \frac{ax^2}{r^3}  & \frac{xy}{r^3} - \frac{ayx}{r^3}  &  - \frac{ax}{r^2}        \\
\frac{yx}{r^3} - \frac{ayx}{r^3} & \frac{b}{r} + \frac{y^2}{r^3} - \frac{ay^2}{r^3}  &  - \frac{ay}{r^2}         \\
- \frac{ax}{r^2} & - \frac{ay}{r^2}  & \frac{b}{r} 
\end{array}\right]$
\end{footnotesize}
\vspace{0.2 cm}
\end{center}

\subsection{Seismic signal on the seismometers}

The ground motion is generated by the three lander feet and is felt by the seismometer through the three SEIS feet.  There are two components to the acceleration felt by SEIS: the acceleration from the direct motion of the ground, and the acceleration due to different vertical displacements of the SEIS feet that causes an inclination of the seismometer in the gravity field (\fig{SEISTiltNEW}). 

\figuremacroW{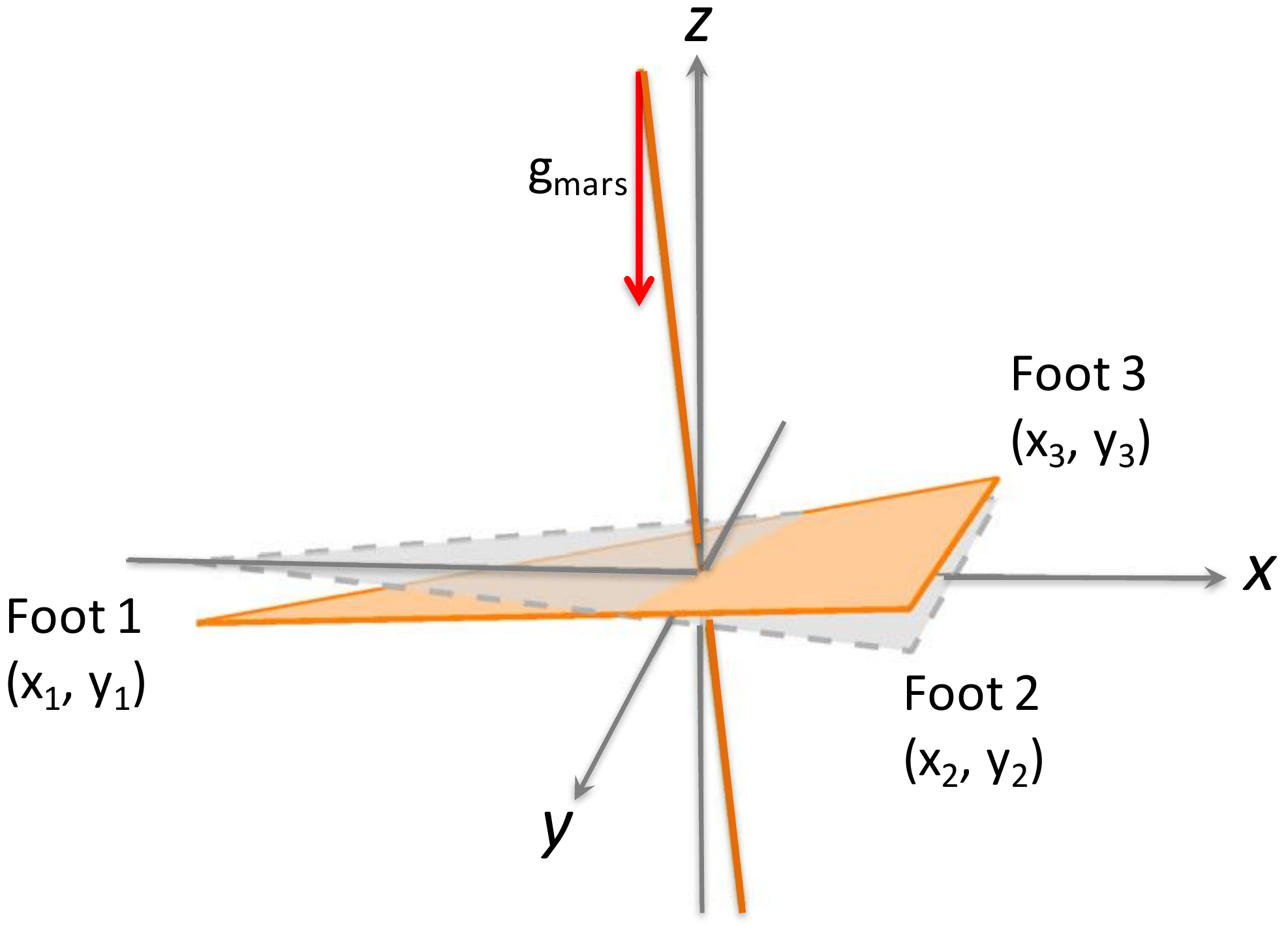}{Tilt noise schematic}{Schematic explaining the tilt noise seen by the seismometer due to the different vertical displacements of the three feet in the gravity field. The reference, perfectly horizontal, seismometer is shown in light grey, and the tilted seismometer is shown in orange. When tilted, the horizontal seismometer axes measure a component of the Martian gravity.}{0.8}

Assuming that the tilt is small, the magnitude of the acceleration due to the tilt in the two horizontal axes ($Ax_{tilt}$ and $Ay_{tilt}$) can be approximated by:

\begin{align}
Ax_{tilt} & = g_{mars} |\frac{(\Delta z_2 +  \Delta z_3)/2 - \Delta z_1}{x_1 - x_2}|\\
Ay_{tilt} & = g_{mars} |\frac{\Delta z_2 -  \Delta z_3}{y_3- y_2}|\\
\end{align}

where $\Delta z_1$, $\Delta z_2$  and $\Delta z_3$ are the vertical displacements of the ground under SEIS feet 1, 2 and 3, respectively, $x_1$, $x_2$ are the $x$ coordinates of the feet 1 and 2, $y_2$, $y_3$ are the $y$ coordinates of the feet 2 and 3.

The total displacement, and thus acceleration, of SEIS due to the direct ground motion is given by the mean displacement of the ground under the three SEIS feet.   The acceleration from the direct motion of the ground is larger at higher frequencies and is the only contribution on the vertical ($z$) axis.  The tilt noise is the dominating noise contribution on the horizontal axes at low frequency.  The total noise that will be registered on the $x$ and $y$ axes is then the sum of the two components of acceleration.  

The resulting day and night time horizontal and vertical noise levels for the baseline configuration (\fig{landerdeploy}) are given in \fig{LanderNoiseExampleNEW}. The influence of the different wind speed spectrum amplitudes is demonstrated in the figure and the complete list of parameters used is provided in \tbl{parameters}. It can be seen that the vertical noise is never expected to exceed the system level noise requirement, and the horizontal noise is only expected to exceed the system level noise requirement for the 95\% wind profile during the day time.

\figuremacroW{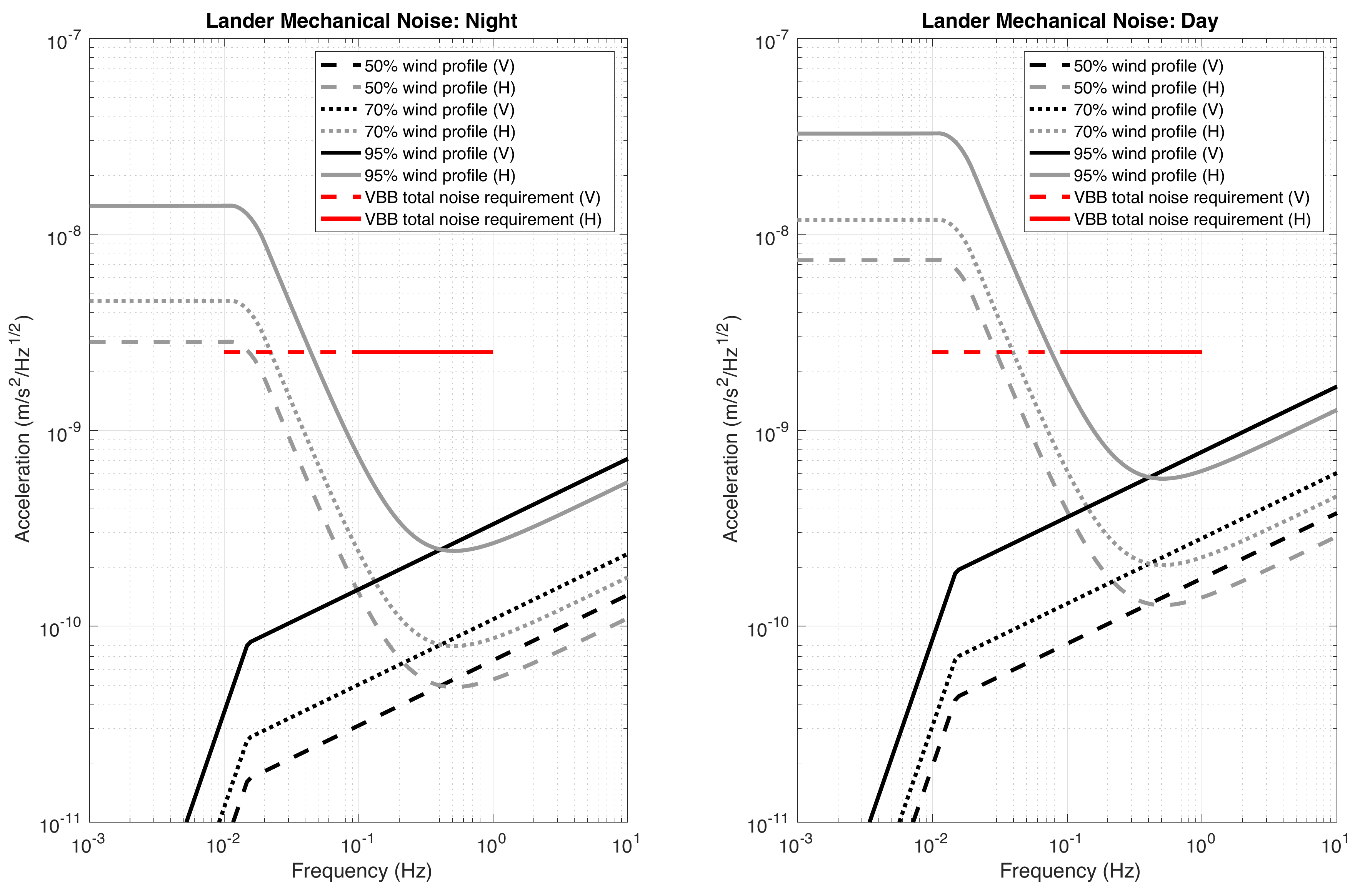}{Lander mechanical noise for the baseline SEIS deployment configuration}{The (left) night time and (right) day time noise on the horizontal (light grey) and vertical (black) axes due to the mechanical noise produced by the wind on the lander.  Rather than show both the acceleration in $x$ and $y$, the horizontal noise is the largest of the two contributors. The different lines show the predicted noise for the 50\% wind profile, the 70\% wind profile results and the 95\% wind profile results.  Also shown are the total VBB horizontal and vertical noise requirements (solid red and dashed red lines, respectively). These simulations assume the baseline parameters given in \tbl{parameters} including that the wind is coming exactly from a North-West direction.}{1}

\begin{table}[htbp!]\footnotesize
\caption{Complete list of parameters used in baseline lander mechanical noise simulations}
\begin{tabularx}{\textwidth}{lc}
 \hline
 \bf{Parameter} & \bf{Value} \\  
 \hline
Lander mass, $m$ & 365 kg \\
Lander height (height of solar panels) & 1.07 m \\
Lander CoG coordinates (in lander centered frame) & (-0.038, 0.001, 0.777) m \\
Lander CoP height (in lander centered frame) & 0.777 m \\
Solar panel offset wrt geometric center of lander, $x_o$ & -0.49 m \\
Solar panel diameter/chord length, $\lambda$ & 2.165 m\\
Number of lander feet, $N$ & 3 \\
Radius of lander feet, $r$ & 0.145 m \\
Surface area of lander exposed to the lift and drag forces, $S$ &  7.53 m$^2$ \\
Distance from lander center to SEIS centre, along ground & 2.59 m \\
Lander feet coordinates (in lander centered frame) &  (-0.15, 0, 0) m\\
 									& (0.075, -0.13, 0) m \\
 									& (0.075, 0.13, 0) m \\
SEIS feet coordinates (in SEIS centered frame) &  (-0.15, 0, 0) m\\
 									& (0.075, -0.13, 0) m \\
 									& (0.075, 0.13, 0) m \\
Mars surface gravity, $g$ & 3.71 ms$^{-2}$ \\
Air density, $\rho$ & 2.2e-2 kg m$^{-3}$ (night);\\
& 1.55e-2 kg m$^{-3}$ (day) \\
Most probable wind direction & from the N-W ($\beta$ = 45$^\circ$) \\
Worst case vertical angle of attack of the wind, $\alpha$ & 15$^\circ$ \\
Surface roughness, $z_0$ & 0.01 m \\
$P$-wave velocity in regolith at reference pressure, $v_{P_{ref}}$ & 265 ms$^{-1}$ \\
$S$-wave velocity in regolith at reference pressure, $v_{S_{ref}}$ & 150 ms$^{-1}$ \\
Bulk density of regolith at reference pressure, $\rho_r$ & 1665 kg m$^{-3}$ \\
Reference pressure, p$_{ref}$ & 25 kPa \\
Reference height for wind calculations, $z_{r}$ & 1.61 m \\
\hline
\end{tabularx}
\label{t:parameters}
\end{table}

\section{Noise maps and Insight deployment zone considerations}\label{s:maps}

Noise ``maps'' have been developed to indicate where the highest and lowest mechanical noise levels are expected to be found within the SEIS deployment zone (\fig{landerdeploy}), for a given set of regolith and wind properties.  Once on Mars, these noise maps will be updated to account for the in-situ parameters and will contribute to the SEIS site selection. 

As an example, \fig{NoiseMaps} shows five noise maps corresponding to the lander mechanical noise at different frequencies (0.01 Hz, 0.1 Hz and 1 Hz) and on different axes (horizontal or vertical). In these calculations it is assumed that the wind is coming from the baseline direction (from the NW) and the 70\% day wind profile is used. The remaining parameters are given in \tbl{parameters}. Each image covers an area of 7 m $\times$ 7 m, centred on the geometric centre of the lander. The colour code in the maps indicates the noise level with respect to the noise budget allocation for the lander mechanical noise \citep[\tbl{NoiseBudget}; for more information see][]{mimoun2016}. 

\begin{table}[htbp!]\scriptsize
\caption{SEIS lander mechanical noise budget}
\begin{tabularx}{\textwidth}{ cc | ccc }
\hline
 \multicolumn{2}{c|}{\bf{Horizontal}}  & \multicolumn{3}{c}{\bf{Vertical}}\\  
 \hline
 \bf{0.1 Hz} & \bf{1 Hz} & \bf{0.01 Hz} & \bf{0.1 Hz} & \bf{1 Hz} \\
1e-9 ms$^{-2}$Hz$^{-1/2}$ & 5e-10 ms$^{-2}$Hz$^{-1/2}$& 5e-10 ms$^{-2}$Hz$^{-1/2}$& 5e-10 ms$^{-2}$Hz$^{-1/2}$ & 1e-9 ms$^{-2}$Hz$^{-1/2}$  \\ 
\hline
\end{tabularx}
\label{t:NoiseBudget}
\end{table}

\figuremacroFive{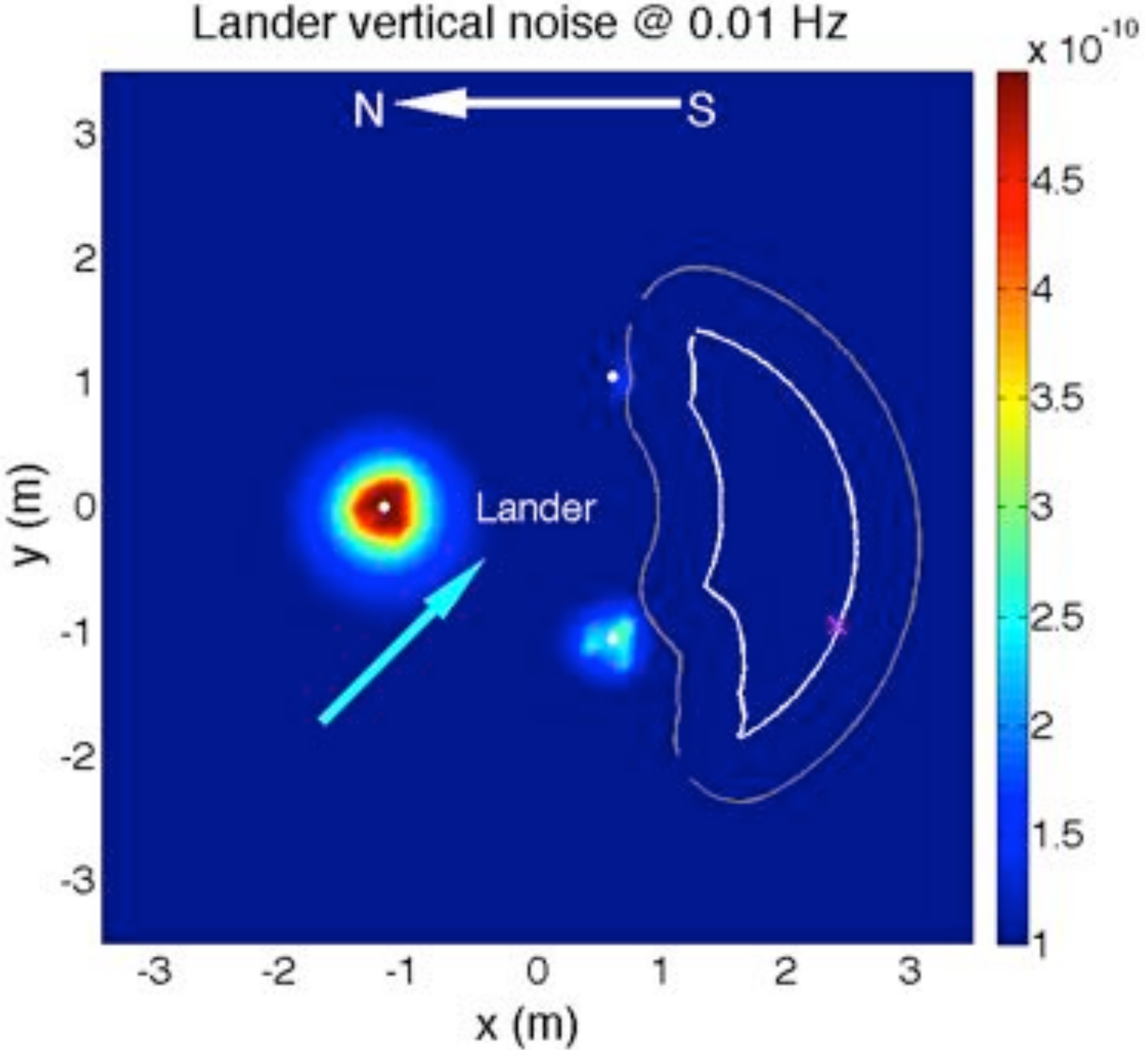}{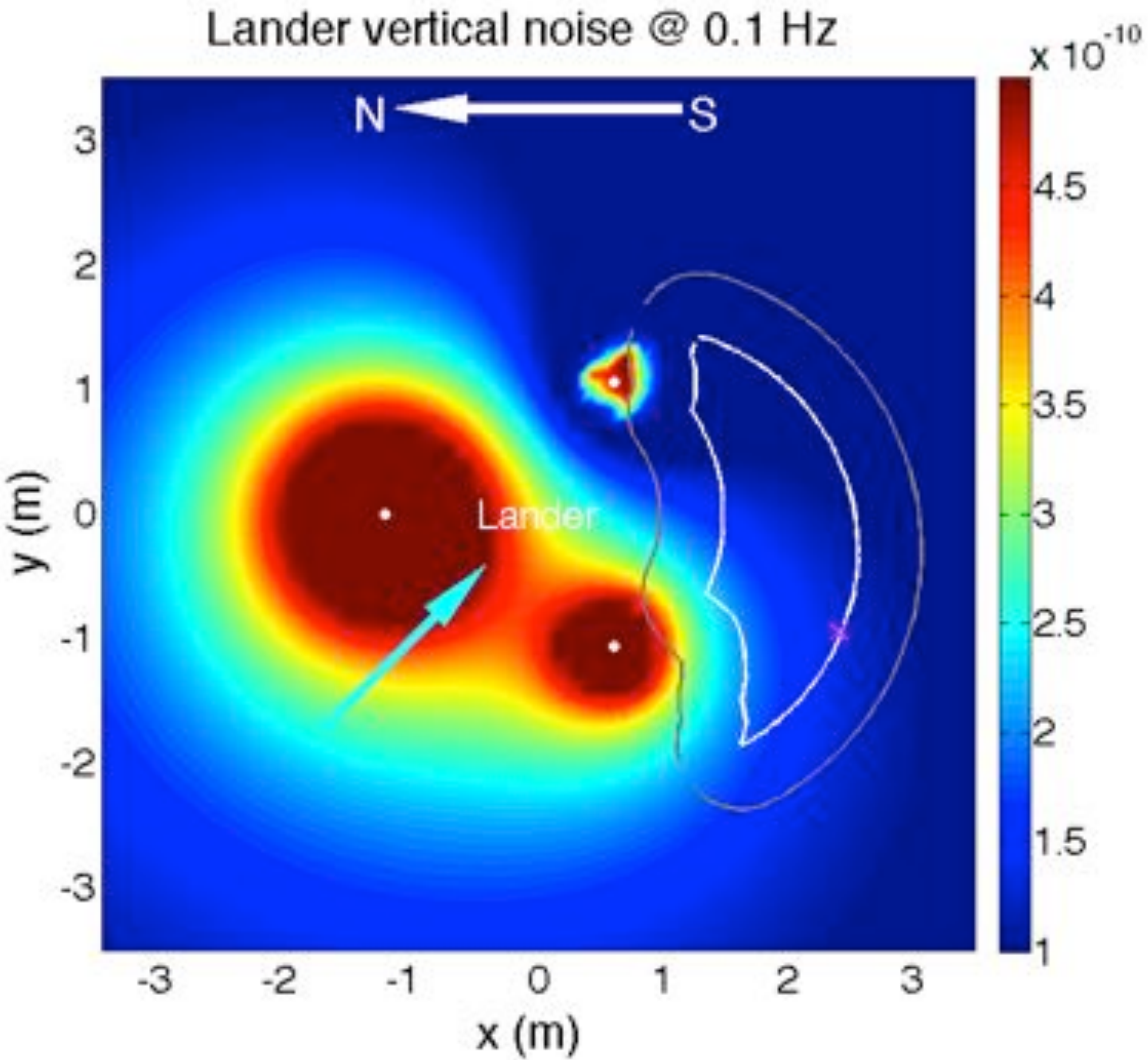}{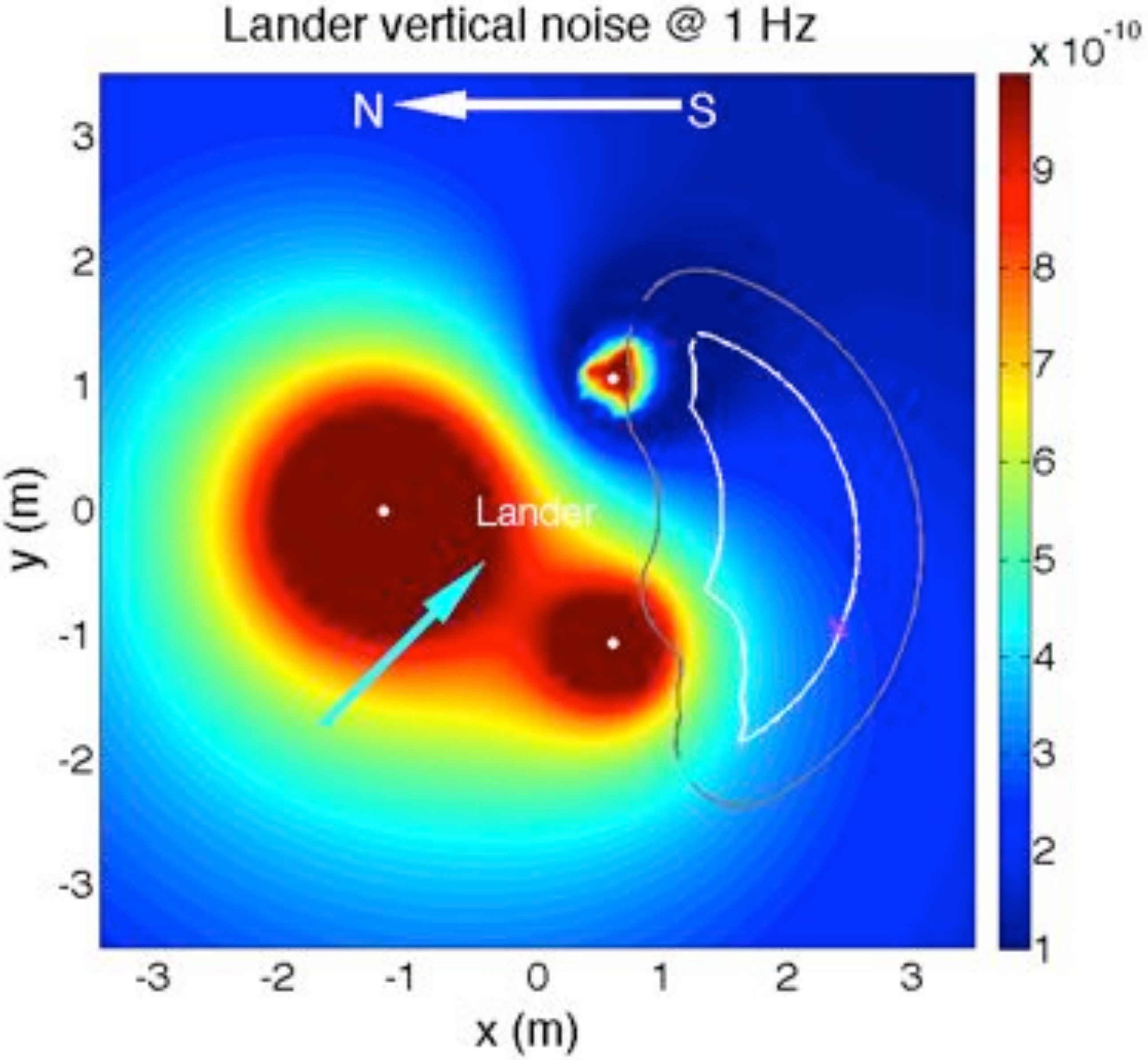}{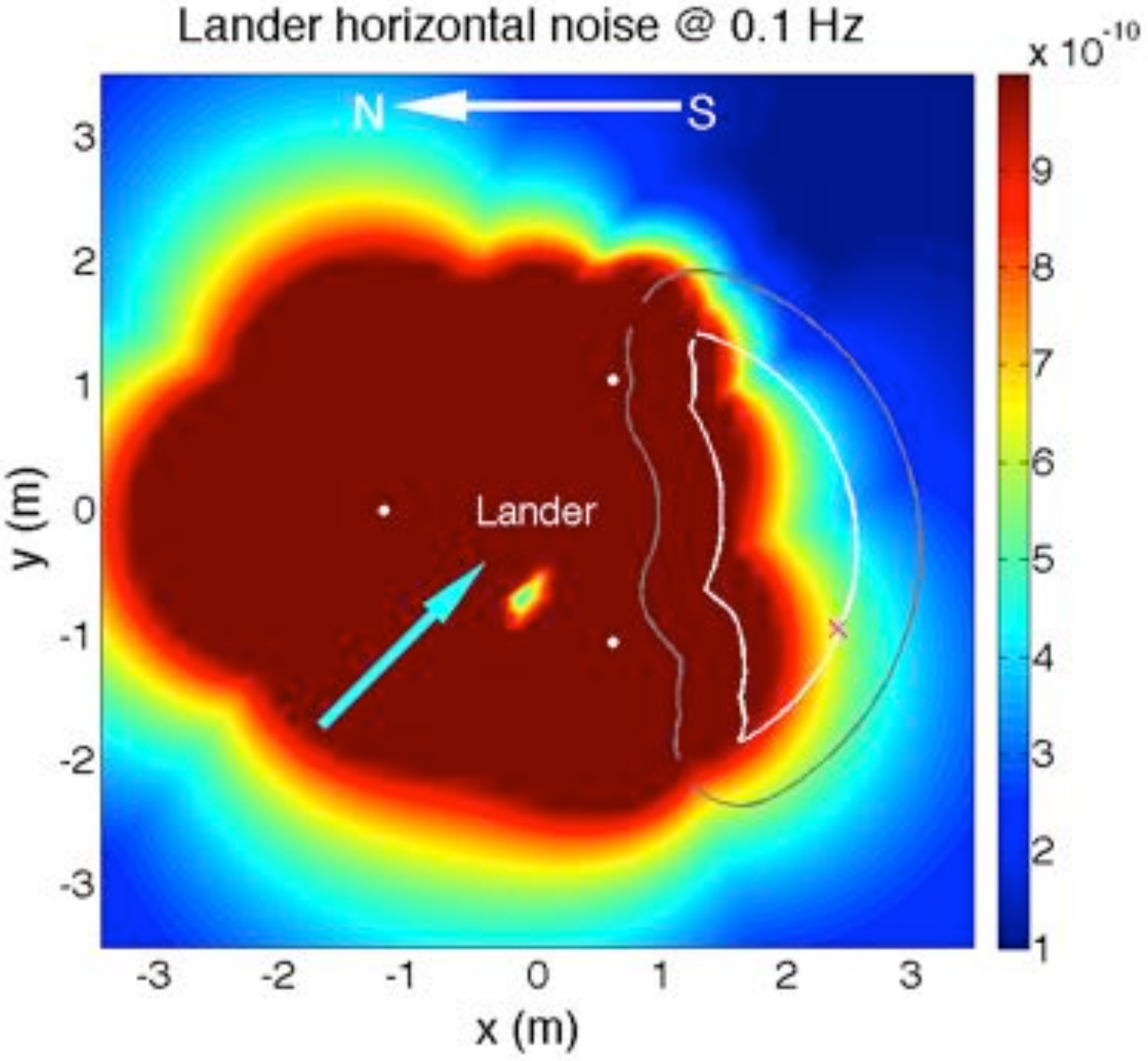}{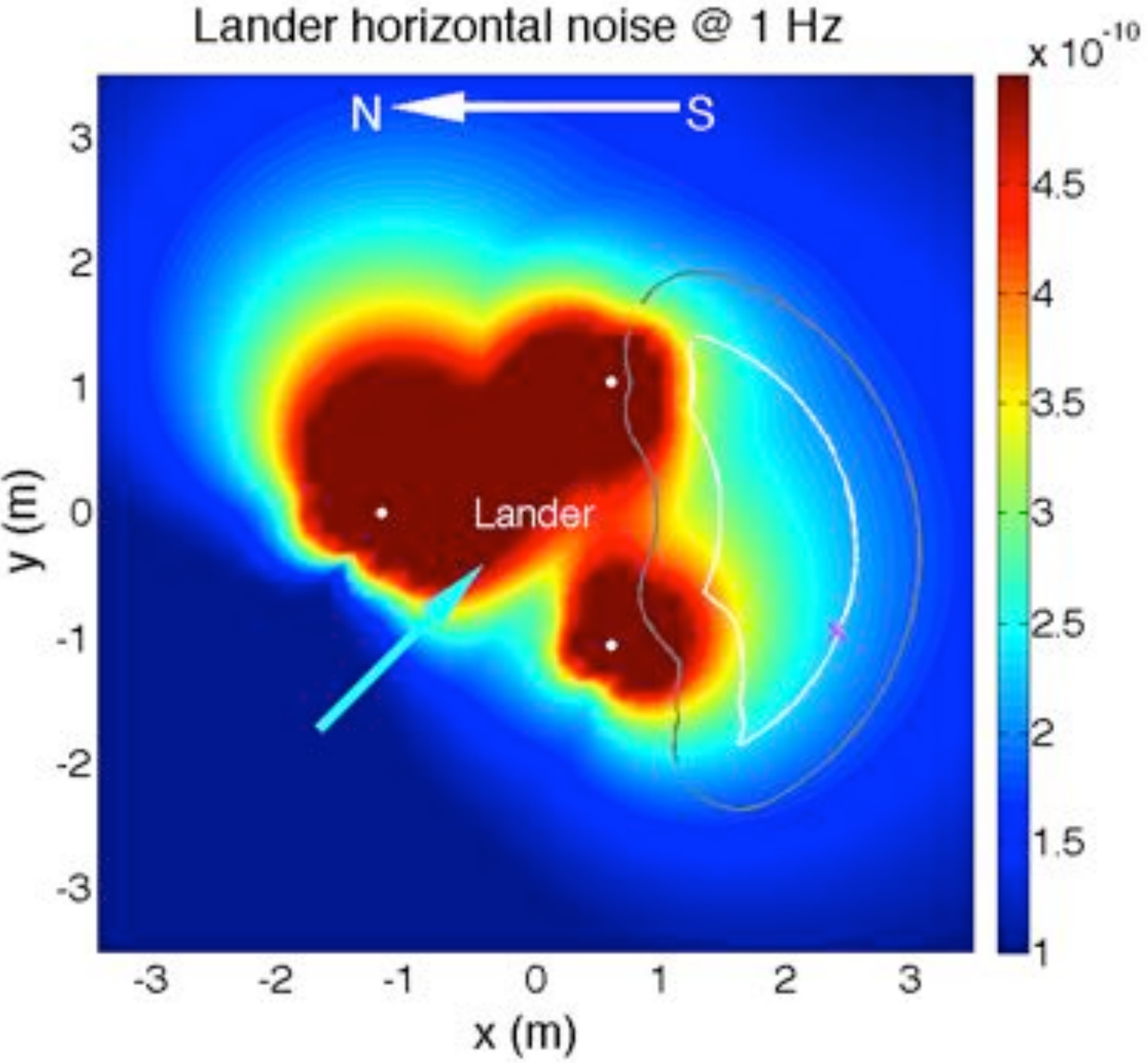}{NoiseMaps}{Example lander mechanical noise maps}{Vertical lander mechanical noise maps at (a) 0.01 Hz, (b) 0.1 Hz and (c) 1 Hz and horizontal mechanical noise maps at (d) 0.1 Hz and (e) 1 Hz. The units of the colour bars are ms$^{-2}$Hz$^{-1/2}$. The colour code indicates the noise level with respect to the noise budget allocation for the lander mechanical noise (\tbl{NoiseBudget}), dark blue being far below the noise budget allocation and dark red is at, or above, the noise budget allocation.  The three lander feet are indicated by the white circles, the possible SEIS and Wind and Thermal Shield deployment zones are indicated by the white and grey outlines, respectively. The SEIS baseline deployment location is indicated by the magenta cross. The wind direction in from the NW as indicated by the cyan arrow and the 70\% day wind profile is used. The remaining parameters are given in \tbl{parameters}. Each image covers an area of 7 m $\times$ 7 m, centred on the geometric centre of the lander.}

\section{Sensitivity of the results to key environment parameters}
 
In addition to depending on the location of SEIS with respect to the lander and the amplitude of the wind speed squared spectrum, the noise that will be registered on SEIS will depend strongly on the incoming wind direction, the amplitude of the wind speed squared spectrum linear model and the ground properties.  A softer ground (lower $v_P$, $v_S$ and $\rho_r$), for example, will deform more easily and thus the low frequency tilt noise will be increased.  Here we investigate the sensitivity of our results to these key parameters by means of a Monte Carlo analysis.

In total 100,000 different parameter combinations were tested and the results of the Monte Carlo analysis are expressed as a cumulative distribution function in \fig{MonteCarloResults_NewNoFcut}. The incoming horizontal wind direction is selected randomly from the interval of 0$^\circ$ to 359$^\circ$. Due to the lack of in-situ data and limited experimental data, the regolith ground properties ($\rho_r$, $v_P$, $v_S$) are selected randomly from a uniform distribution over the interval of -3$\sigma$ to 3$\sigma$, rather than from a normal distribution. Similarly, the surface roughness is selected randomly from the range 1 mm to 5 cm. The amplitude ($B$) of the day and night wind speed squared spectra follows the lognormal distributions presented in \fig{PDF}. We assume that SEIS is in the baseline deployment position (\fig{landerdeploy}) and for the remaining parameters, the baseline values are used (\tbl{parameters}).  

Based on our current assumptions, in the baseline deployment configuration, the lander mechanical noise level is expected to always be below the total VBB noise requirement except for the horizontal noise at 0.1 Hz,  and the vertical noise at 1 Hz, which very occasionally ($\le$3\% and $\sim$0.5\% of the time, respectively) exceed the requirement of 2.5 $\times$ 10$^{-9}$ ms$^{-2}$ Hz$^{-1/2}$ (see lower plots of \fig{MonteCarloResults_NewNoFcut}). The lander mechanical noise is, therefore, not expected to endanger the InSight mission goals.

\figuremacroW{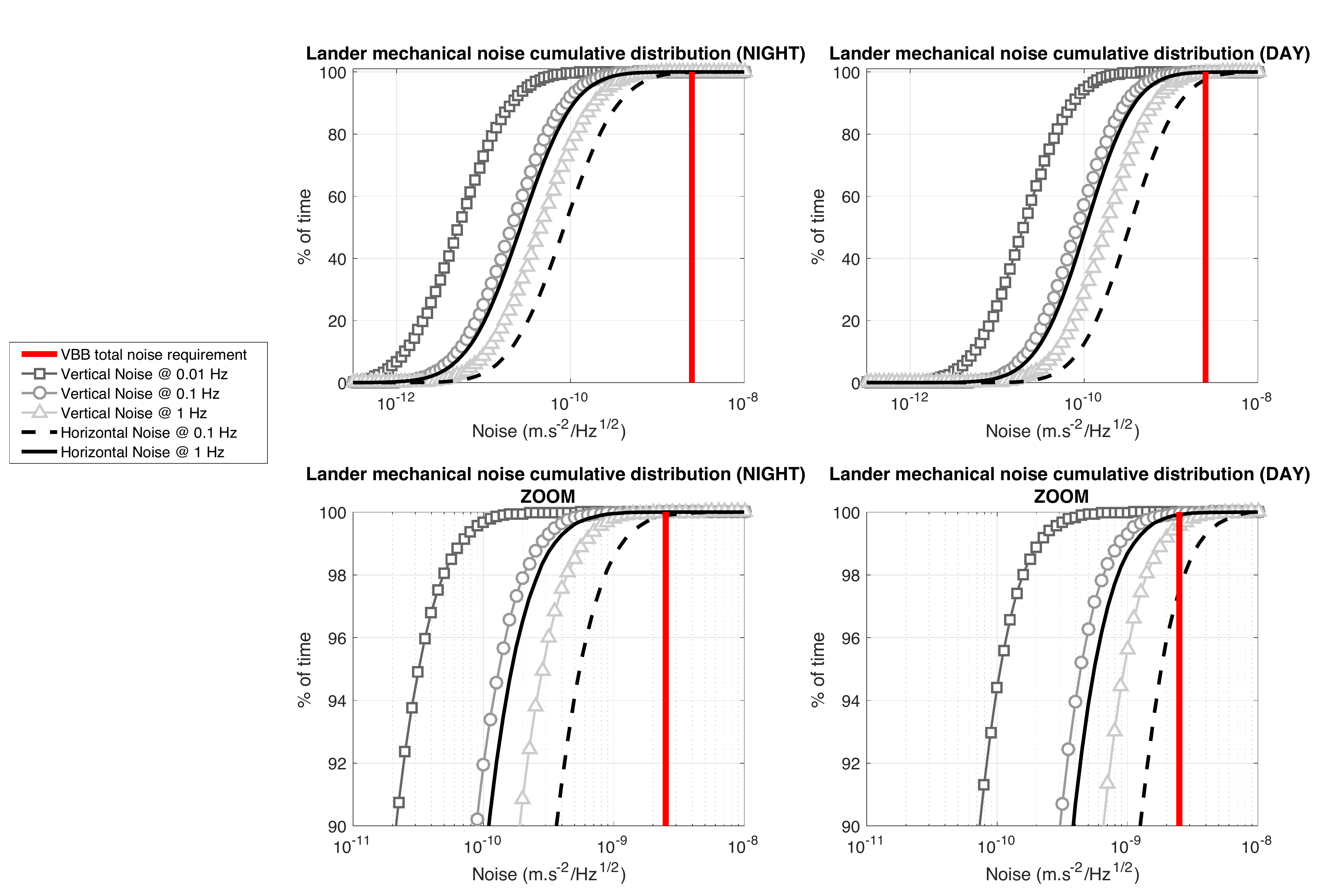}{Monte Carlo simulations results}{The (left) night time and (right) day time cumulative distribution functions for the predicted lander mechanical noise registered on SEIS. The lower figures have a reduced range, showing the percentage of time when the lander mechanical noise is expected to be above the VBB total noise requirements (solid red line).}{1}

\newpage
\section{Influence of the solar panel resonances}
\label{s:resonances}

In order to estimate the impact of the solar panel resonances, we attempt to identify the solar panel flapping mode resonance of the Phoenix lander on Mars in 2008.  The Phoenix Surface Stereo Imager captured an image of the lander deck and solar panels on sol 96 (\fig{SSIpanels}). While the deck and background are steady in the image, the solar panels appear vertically blurred. The amplitude of the motion blur increases from left to right though the solar panel cantilever structure. The analysis of the photometric image reveals two distinct upper and lower ``ghost'' positions were captured in the image. More importantly, the lower ghost position captured is about twice as bright as the upper position \fig{SSIpanels}g. This shows that the image captured more than half a period of the flapping motion, but about less than 3/2 of a period. Given the image exposure time of 0.71 seconds, this sets the frequency boundaries for the solar panel mode: between 0.7 and 2.1 Hz. According to the scale of the components seen on the image, the amplitude of the vertical motion on the solar panel edge is about 3.4 mm.

The transfer function due to the vertical solar panel vibrations can be expressed via the following equation :

\begin{equation}
T_i(p) = \frac{2 \zeta \omega_i p + \omega_i^2}{p^2 + 2 \zeta \omega_i p + \omega_i^2}
\end{equation}

where $\zeta$ is the passive damping ratio assumed to be equal to 0.05, $\omega=2\pi f$,  $p$ is the laplace variable (\ie $i\omega$) and $\omega_i=2\pi f_i$ where $f_i$ is the resonance frequency of the solar panels (assumed to be 1.4 Hz, the mean value of the boundaries determined above). The influence of the solar panel vibrations can be seen as a peak in the seismic signal at the resonant frequency of 1.4 Hz (\fig{Resonances}).  

The detailed simulation of the solar panel vibrations is not within the scope of this paper and will be included in future work. The resonant frequencies of the InSight solar panels are not expected to be identical to those of Phoenix as the InSight solar panels are larger.  Similarly, the passive damping ratio will have to be calibrated using the correct values measured under Martian surface pressure.  

This exercise, nonetheless, demonstrates the type of signal that we may see on SEIS due to the InSight solar panel resonance modes. As the resonance frequency calculated here (1.4 Hz) is above the VBB bandwidth, a similar resonant frequency for the InSight solar panels would not impact the InSight mission goals. 

\figuremacroW{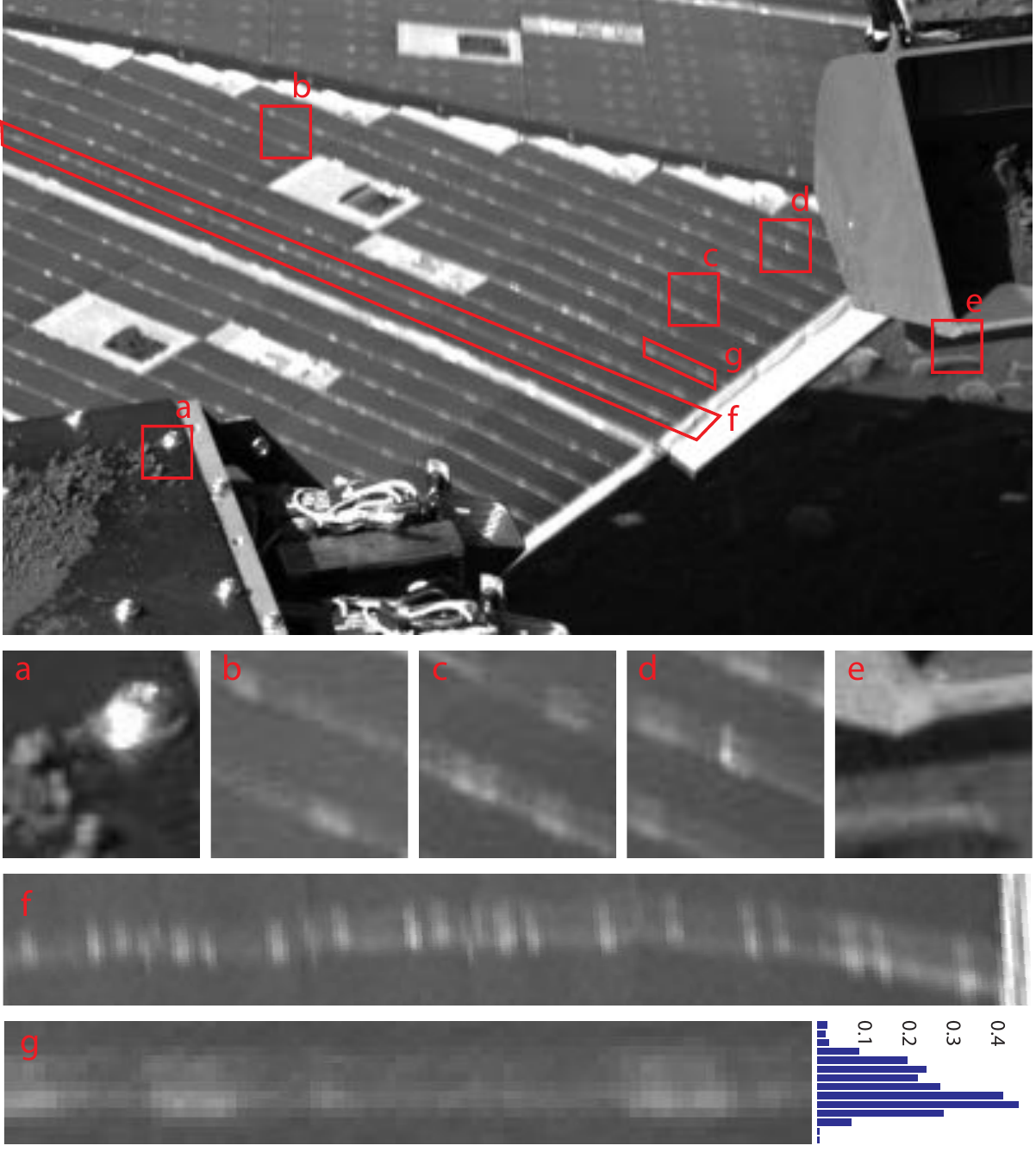}{Phoenix lander solar panel vibrations}{Analysis of the Image taken by the phoenix lander on sol 96 (SS 096 RAL 904723784\_1B300R6 M1 radiometrically corrected). Close-up views highlight steady rover deck (a) and background (e) compared to solar panels (b, c and d). The vertically stretched close-up on a bright cable (f) shows the vertical motion blur increasing from left to right. The pixel rows sum on the close up of a bright solar panel cable (g) shows the difference of brightness for the two ghost positions.}{1}

\figuremacroW{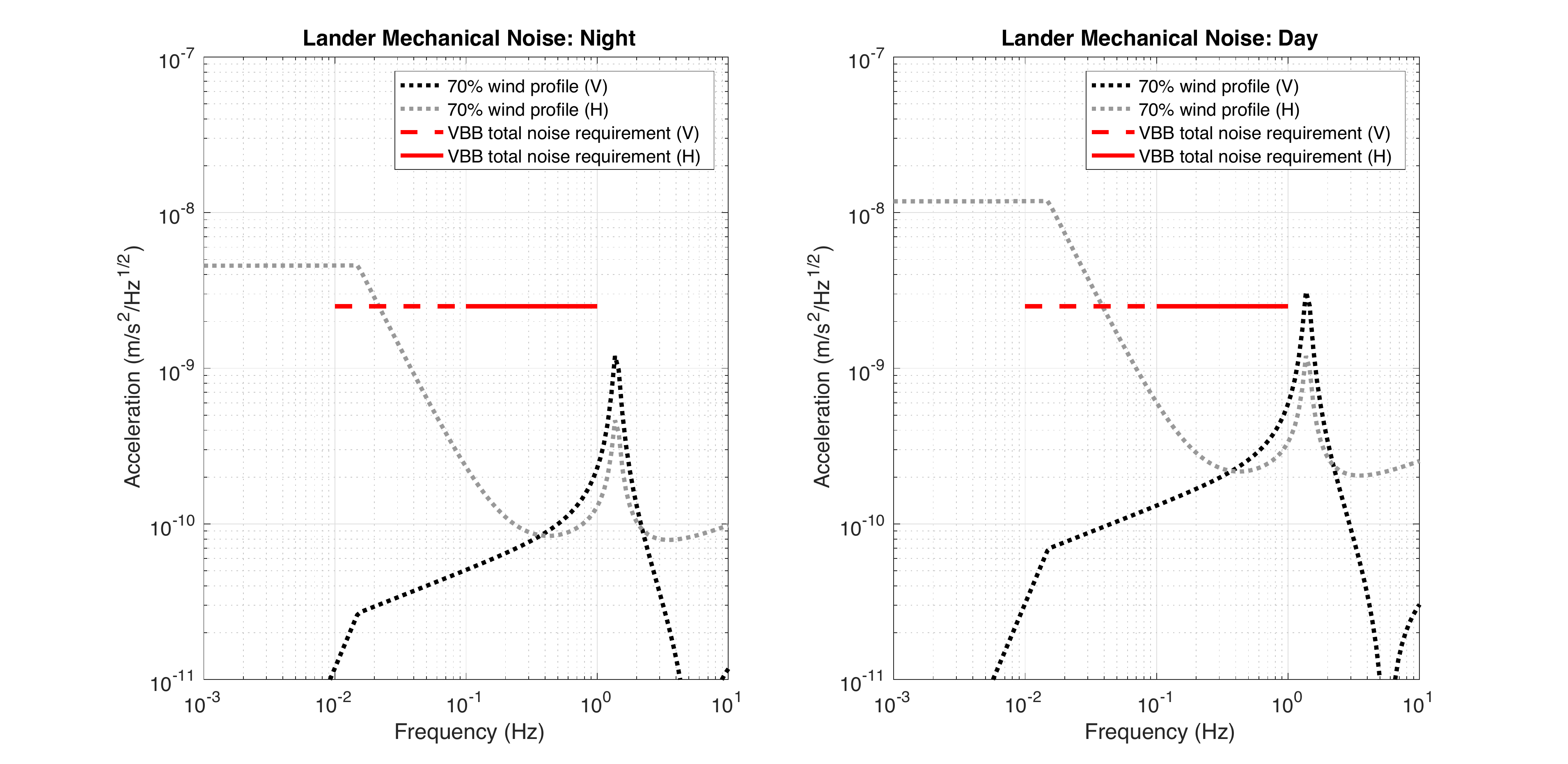}{Lander mechanical noise with solar panel resonances}{The (left) night time and (right) day time noise on the horizontal (grey dotted line) and vertical (black dotted line) axes due to the mechanical noise produced by the wind on the lander assuming the 70\% wind profile.  Rather than show both the acceleration in $x$ and $y$, the horizontal noise is the largest of the two contributors. The long-period noise on the horizontal axes remains an important signal, even with respect to the resonance. This is because the amplitude of the dynamic pressure spectrum is larger at low frequencies, and because, at low frequencies, the horizontal axes are very sensitive to any inclination of the seismometer in the Martian gravity field (\fig{SEISTiltNEW}). Also shown are the total VBB horizontal and vertical noise requirements (solid red and dashed red lines, respectively). These simulations assume the baseline parameters given in \tbl{parameters} and include the solar panel resonance as estimated from the Phoenix lander images - see \sect{resonances}.}{1}

\section{Calculation of the WTS and HP3 mechanical noise}

Our complete mechanical noise model has also been adapted to estimate the wind-induced mechanical noise on SEIS coming from the Wind and Thermal Shield (WTS) and the Heat Flow and Physical Properties Package (HP3).  The WTS and HP3 are assumed to be at a local topographic slope normal to gravity over 1 to 5 m length scales. The wind is, therefore, likely to be parallel to the surface and, thus, at zero angle of attack as far as the WTS and HP3 is concerned. However, as the WTS is a bluff body with a cavity under it that is not exposed to the flow, there should be a vertical lift force even at zero angle of attack because the pressure outside the WTS is modified by the flow over it, compared to the interior gas which, in the limiting case of no leakage under the WTS, should be stagnant. This is not the case for the HP3 which is assumed to experience no lift force. The WTS lift and WTS and HP3 drag forces are calculated using the coefficients and surface areas given \tbl{WTSparams} and \tbl{HP3params}.  Note that the WTS and SEIS feet are assumed to be radially aligned \ie in the `clocked position' (see \fig{landerdeploy}). The calculations of the forces exerted on the ground at the WTS and HP3 feet, the ground deformation and the signal felt by SEIS are performed exactly as for the lander mechanical noise.

The resulting day and night time horizontal and vertical WTS and HP3 noise levels for the baseline configuration are given in \fig{WTSHP3NoiseExample}.  It can be seen that the WTS and HP3 mechanical noise is never expected to exceed the instrument-level noise requirement.

\figuremacroTwo{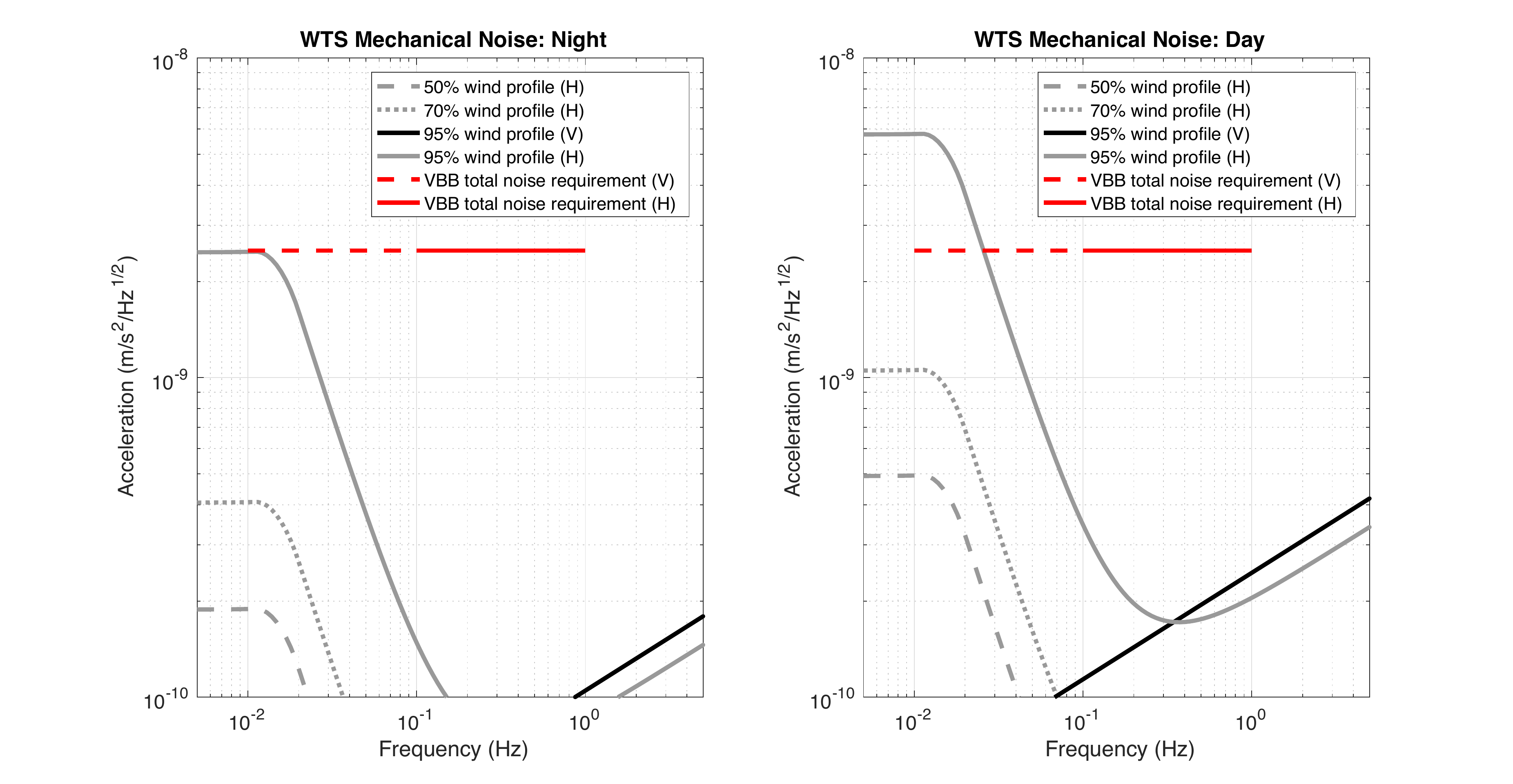}{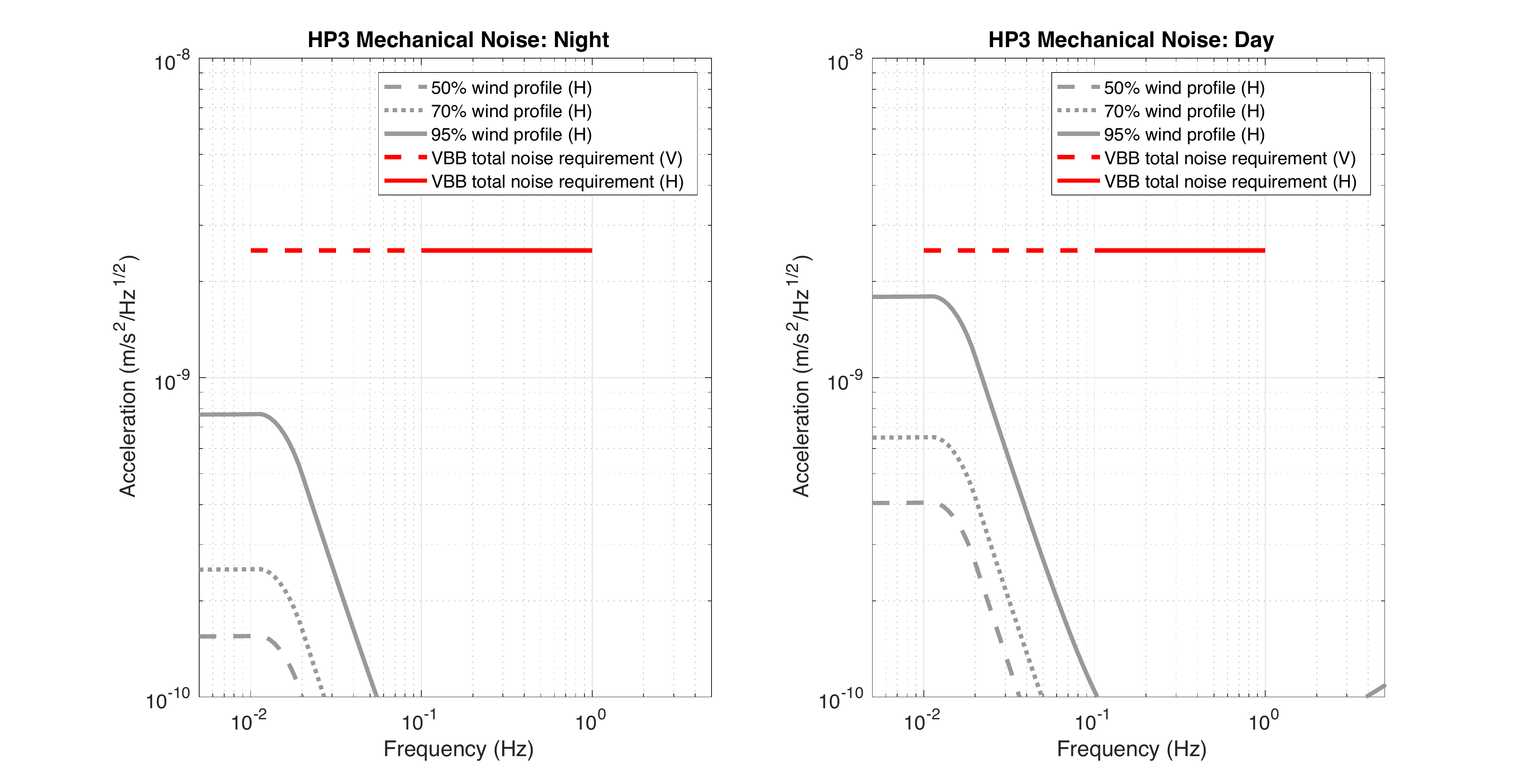}{WTSHP3NoiseExample}{WTS and HP3 mechanical noise for the baseline SEIS deployment configuration}{The (upper left) night time and (upper right) day time noise on the horizontal (light grey) and vertical (black) axes due to the mechanical noise produced by the wind on the WTS. The (lower left) night time and (lower right) day time noise on the horizontal (light grey) and vertical (black) axes due to the mechanical noise produced by the wind on the HP3.   Rather than show both the acceleration in $x$ and $y$, the horizontal noise is the largest of the two contributors. The different lines show the predicted noise for the 50\% wind profile, the 70\% wind profile results and the 95\% wind profile results.  Also shown are the total VBB horizontal and vertical noise requirements (solid red and dashed red lines, respectively). These simulations assume the baseline parameters given in \tbls{parameters}{HP3params} including that the wind is coming exactly from a North-West direction.}

\begin{table}[htbp!]\footnotesize
\caption{Complete list of parameters used in the WTS mechanical noise simulations. Note that the WTS and the SEIS centred frames are co-located. The environment variables remain the same as in \tbl{parameters}.}
\begin{tabularx}{\textwidth}{lc}
 \hline
 \bf{Parameter} & \bf{Value} \\  
 \hline
WTS mass & 9.5 kg \\
WTS height & 0.4 m \\
WTS CoG coordinates (in WTS centered frame) & (0, 0, 0.09) m\\
WTS CoP coordinates (in WTS centered frame) & (0, 0, 0.09) m \\
Number of WTS feet & 3 \\
Radius of WTS feet & 0.04 m \\
Surface area of WTS exposed to the lift and drag forces &  0.21 m$^2$ \\
WTS lift coefficient & 0.36 \\
WTS drag coefficient & 0.45 \\
WTS feet coordinates (in WTS centered frame) &  (-0.46, 0.08, 0) m\\
 									& (-0.23, -0.40, 0) m \\
 									& (0.23, -0.40, 0) m \\								
\hline
\end{tabularx}
\label{t:WTSparams}
\end{table}

\begin{table}[htbp!]\footnotesize
\caption{Complete list of parameters used in the HP3 mechanical noise simulations. The environment variables remain the same as in \tbl{parameters}.}
\begin{tabularx}{\textwidth}{lc}
 \hline
 \bf{Parameter} & \bf{Value} \\  
 \hline
HP3 mass & 2.9 kg \\
HP3 height & 0.43 m \\
HP3 CoG coordinates (in HP3 centered frame) & (0, 0, 0.11) m\\
HP3 CoP coordinates (in HP3 centered frame)& (0, 0, 0.11) m \\
Number of HP3 feet & 4 \\
Radius of HP3 feet & 0.04 m \\
Surface area of HP3 exposed to the lift and drag forces &  0.1 m$^2$ \\
HP3 drag coefficient & 1.2  \\
Distance from HP3 center to SEIS centre, along ground & 1.6 m \\
HP3 feet coordinates (in HP3 centered frame) &  (-0.19, 0.08, 0) m\\
 									& (-0.19, -0.08, 0) m \\
 									& (-0.39, -0.17, 0) m \\
 									& (-0.39, 0.17, 0) m \\
\hline
\end{tabularx}
\label{t:HP3params}
\end{table}

\section{Discussion and conclusions}

We have developed a complete model for simulating the wind-induced lander, WTS and HP3 mechanical noise on the InSight seismometers. The results indicate that, for the baseline SEIS deployment position, the lander mechanical noise will rarely ($<$3\% of the time) exceed the total noise requirement and should, therefore, not prevent InSight from achieving the key mission objectives.  Our mechanical noise model has also been adapted to model the mechanical noise coming from the wind stresses on the wind and thermal shield and the second InSight instrument, HP3 as these will also be transmitted to SEIS via the ground. Again, these noise contributions are not likely to exceed the instrument level noise requirements.

The wind speed squared spectrum is likely to be more complicated than the simplistic form we derived from the available in-situ data and theoretical arguments.   However, it should be noted that, in the past, Martian wind properties have been determined from the Viking Lander seismic measurements. \cite{anderson1977} found that the lander displacement correlates well with wind velocity. In fact, there is a strong agreement that the seismic amplitude is proportional to the square of the wind speed \citep{anderson1977, nakamura1979} confirming at least our hypothesis that the seismic signal should be proportional to the dynamic pressure. 

Another, and perhaps the most important, assumption in our model is that the Martian ground behaves elastically. As the InSight landing site will almost definitely be covered by regolith there is likely to be plastic deformation occurring and thus seismic anelasticity in the regolith that has not been accounted for in our elastic model.  This means that the seismic amplitudes presented in this paper are probably upper bounds and the lander mechanical noise may actually be much lower than we predict as part of the lander forcing will generate regolith flows instead of elastic deformations.

The other assumption of an homogeneous elastic half-space does not take into account the fact that the Martian subsurface is likely to be layered. Practically, this means that a larger rigidity at long periods must be used, as the long period signals will be sensitive to the deeper structure. The layered subsurface may also lead to seismic waves being reflected several times at the layer boundaries resulting in resonances in the regolith, especially at high frequencies. Such waves and dynamic noise are neglected in our homogeneous elastic half-space assumption.  However, reflections are only expected to arise at  frequencies much larger than 1 Hz. For example, for 5 meter thick regolith layer with 150 ms$^{-1}$  shear waves on bed rock, a first resonance might occur for about one fourth of wavelength (\ie at about 7.5 Hz). Future work will include studying the anelastic effects that may be expected in the Martian regolith \citep{Teanby2016}, and the impact of a layered sub-surface \citep{kenda2016}

This paper currently concentrates on the  [0.01-1 Hz] bandwidth as the very broad band seismometer is the critical instrument for achieving the Insight mission objectives. However, as we have seen in \sect{resonances}, the lander resonances may significantly increase the lander mechanical noise at higher frequencies and, therefore, could also impact the short-period seismometer. Studying the lander resonances in more detail are part of future planned work.

In the future, to improve the accuracy of the noise estimations, more detailed Computational Fluid Dynamics simulations could be performed to model the wind-lander interactions \citep[\eg][]{Chiodini2104,gendron2010}. However, the atmospheric boundary layer is itself fully and highly turbulent, and the lander induced eddies are just one component of the full spectrum that the boundary layer itself presents. 

A detailed analysis of the Mars Science Lab (MSL, Curiosity) Rover Environmental Monitoring Station (REMS) data may also further refine our wind hypotheses. We hope that the current and upcoming space missions (including InSight) will provide more valuable environment data that will lead to a better understanding of the Martian atmosphere and allow our environment models to be improved. Finally, we note that the lander mechanical noise may actually provide an additional seismic source for determining the seismic properties of the Martian subsurface. Future work will include trying to solve the inverse-problem once InSight has arrived on Mars. 

\section{Acknowledgements}

We acknowledge many stimulating discussions related to this subject with members of the SEIS and InSight teams.  The Phoenix Telltale experiment data were obtained from the Planetary Data System.  We are grateful to J. Murphy and J. Tillman for providing the Viking Lander wind data and we thank C. Wilson and A. Spiga for their very useful insights related to the wind properties on Mars and the interpretation of previous in-situ Martian data.  We would also like to thank N. Teanby for his helpful comments on the manuscript. This work has been supported by CNES, including post-doctoral support provided to N. Murdoch. 

\bibliographystyle{aps-nameyear}      

\end{document}